%% file: main.tex
\documentclass[reprint,superscriptaddress,amsmath,amssymb,aps,floatfix,longbibliography]{revtex4-2}
\input{preamble}
\begin{document}
\title{Learning materials spectra across modalities and scales with structure-conditioned neural fields}
\input{sections/authors}
\begin{abstract}
\input{sections/abstract}
\end{abstract}

\maketitle
\input{sections/introduction}

\section{Results}
\input{sections/results_fig1}
\input{sections/results_fig2}
\input{sections/results_fig3}
\input{sections/results_fig4}
\input{sections/discussion}
\input{methods_revision}

\begingroup\small
\input{sections/declarations}
\endgroup
\renewcommand{\bibfont}{\fontsize{9}{11}\selectfont}
\bibliography{references}

\clearpage
\onecolumngrid
\input{si_revision}

\end{document}

%% file: preamble.tex
\usepackage[T1]{fontenc}
\usepackage[utf8]{inputenc}
\usepackage{array}[=2016-10-06]
\usepackage{amsfonts,bm,graphicx,booktabs,tabularx}
\usepackage{xcolor}
\usepackage{xurl}
\usepackage{hyperref}
\hypersetup{colorlinks=true,linkcolor=blue!45!black,citecolor=blue!45!black,urlcolor=blue!45!black,pdfauthor={Tianyu Zhang, Yuan Ni, Roy Liu, Garrett Granroth, Daniel M. Pajerowski, Joshua J. Turner, Yongqiang Cheng, Zhantao Chen},pdftitle={Learning materials spectra across modalities and scales with structure-conditioned neural fields}}
\setcitestyle{super}
\newcolumntype{Y}{>{\raggedright\arraybackslash}X}
\newif\ifdraftnotes
\draftnotesfalse

\newcommand{\mainfig}[2]{%
  \begin{figure*}[!tp]
  \centering
  \ifnum#1=4\relax
  \includegraphics[width=\textwidth,height=#2\textheight,keepaspectratio]{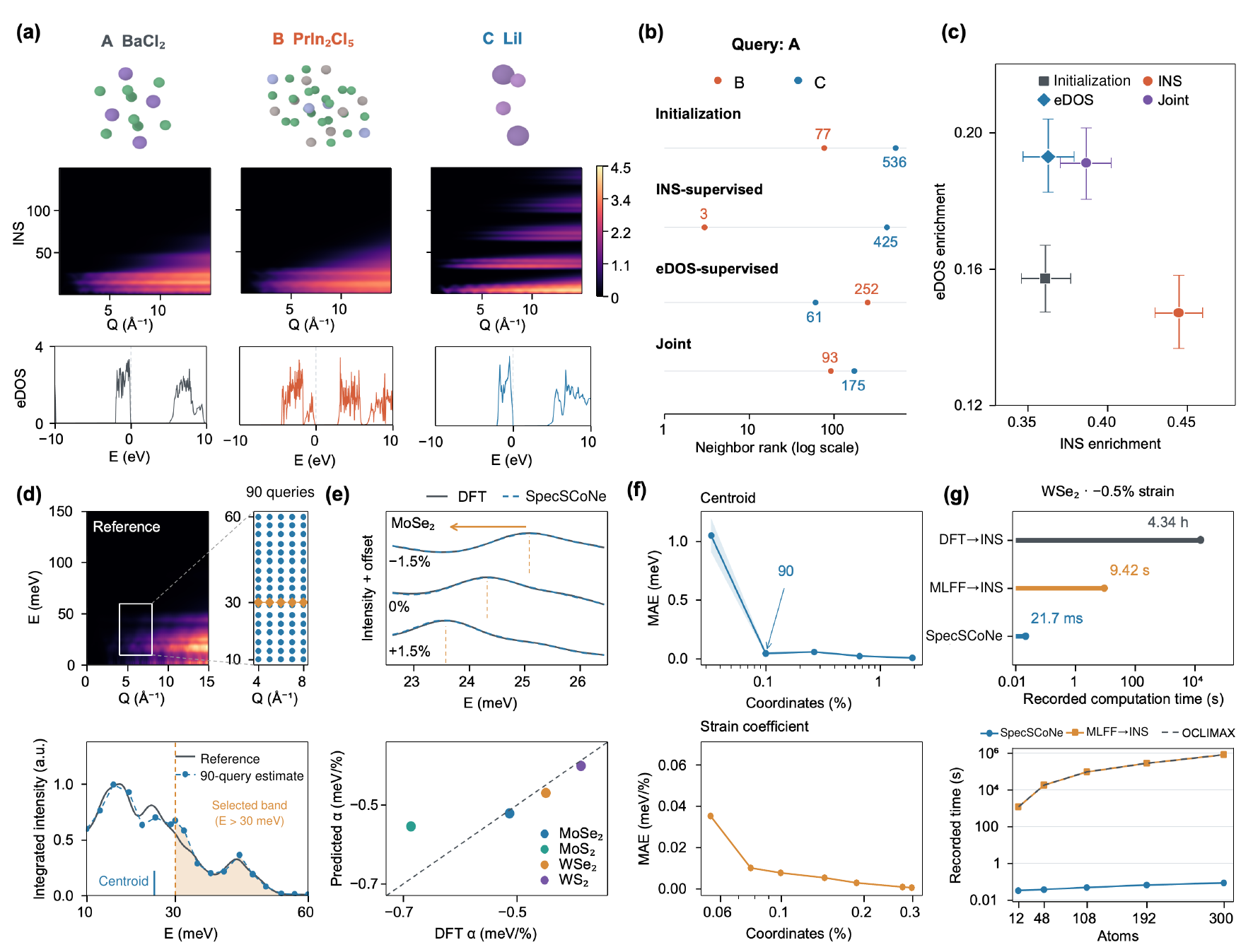}
  \else
  \includegraphics[width=\textwidth,height=#2\textheight,keepaspectratio]{figures/Figure#1.png}
  \fi
  \input{captions/figure#1}\label{fig:#1}
  \end{figure*}}

%% file: sections/authors.tex
\author{Tianyu Zhang}
\affiliation{Group for AI in Materials Modeling and Analysis, The University of Texas at Austin, Austin, TX, USA.}
\affiliation{Texas Materials Institute, The University of Texas at Austin, Austin, TX, USA.}

\author{Yuan Ni}
\affiliation{Linac Coherent Light Source, SLAC National Accelerator Laboratory, Menlo Park, CA, USA.}
\affiliation{Stanford Institute for Materials and Energy Sciences, Stanford University, Stanford, CA, USA.}

\author{Roy Liu}
\affiliation{Group for AI in Materials Modeling and Analysis, The University of Texas at Austin, Austin, TX, USA.}
\affiliation{Department of Computer Science, The University of Texas at Austin, Austin, TX, USA.}

\author{Garrett Granroth}
\affiliation{Neutron Scattering Division, Oak Ridge National Laboratory, Oak Ridge, TN, USA.}

\author{Daniel~M.~Pajerowski}
\affiliation{Neutron Scattering Division, Oak Ridge National Laboratory, Oak Ridge, TN, USA.}

\author{Joshua J.~Turner}
\affiliation{Linac Coherent Light Source, SLAC National Accelerator Laboratory, Menlo Park, CA, USA.}
\affiliation{Stanford Institute for Materials and Energy Sciences, Stanford University, Stanford, CA, USA.}

\author{Yongqiang Cheng}
\email[Corresponding author: ]{chengy@ornl.gov}
\affiliation{Neutron Scattering Division, Oak Ridge National Laboratory, Oak Ridge, TN, USA.}

\author{Zhantao Chen}
\email[Corresponding author: ]{zhantao@austin.utexas.edu}
\affiliation{Group for AI in Materials Modeling and Analysis, The University of Texas at Austin, Austin, TX, USA.}
\affiliation{Texas Materials Institute, The University of Texas at Austin, Austin, TX, USA.}
\affiliation{Walker Department of Mechanical Engineering, The University of Texas at Austin, Austin, TX, USA.}

%% file: sections/abstract.tex

Spectroscopic observables encode rich information about material behavior across diverse, often high-dimensional domains of energy, momentum, and other experimental coordinates, providing critical insights into the physical properties underlying material functionalities and applications. However, learning these responses directly from atomic structures remains challenging due to the complexity of both structural configurations and spectroscopic observation spaces. Here, we develop a structure-conditioned neural-field framework that learns continuous spectroscopic response functions by coupling an equivariant structural encoder with a coordinate-conditioned network. The framework accurately predicts spectroscopic observables across diverse observation domains, ranging from one-dimensional densities of states to multidimensional scattering functions. We show that fine-tuning with limited spectral supervision captures strain- and composition-dependent spectral changes, with validated transfer to alloy configurations containing hundreds of atoms and computational scalability demonstrated for configurations exceeding $10^4$ atoms. Our work establishes a versatile and computationally scalable approach to learning structure-spectroscopy relationships, providing a foundation for spectroscopic analysis and future materials characterization, inverse problems, and experimental design.

%% file: sections/introduction.tex
\section{Introduction}

\newcommand{\ac}[0]{\textcolor{red}{[cite]}}


Spectroscopy probes material dynamics and excitations through responses resolved in experimental coordinates such as energy and momentum transfer~\cite{10.1093/acprof:oso/9780199228676.001.0001,sobota2021arpes,mitrano2024exploring}. These spectra provide insights into electronic, vibrational, and optical properties that depend on the underlying atomic composition and arrangement~\cite{basov2011electrodynamics,fultz2010vibrational}. Understanding the structural origins of these properties and tailoring atomic arrangements for desired applications therefore requires establishing predictive relationships between atomic structures and spectroscopic responses~\cite{chen2021machine,han2025vibrations}. Calculating spectroscopic responses from atomic structures often requires computationally expensive simulations.
Machine learning (ML) is playing an increasingly important role in accelerating computational materials science and structure-property relationship learning~\cite{choudhary2021atomistic,deringer2021gpr} with examples including ML-accelerated molecular dynamics~\cite{unke2021forcefields,chen2022m3gnet,deng2023chgnet,musaelian2023allegro} and surrogate models for density functional theory (DFT) Hamiltonians and electronic structure~\cite{li2022deeph,gong2023deehe3,fiedler2023electronic}. 
However, even with ML-accelerated simulations, predicting spectroscopic responses often requires computationally expensive intermediate calculations and observable-specific post-processing, such as converting interatomic force constants into phonon modes and neutron-scattering intensities~\cite{baroni2001dfpt,togo2023phonopy,fair2022euphonic,cheng2019oclimax,lindgren2025neutron,han2025insbenchmark}. 
This has motivated the development of models that directly predict spectroscopic observables from atomic structures, bypassing these computational bottlenecks~\cite{kong2022mat2spec,cheng2023direct,hung2024optical,carbone2020xas,grunert2024dielectric,lin2026xanes,han2024inspired}. 
Despite substantial progress in predictive accuracy and the range of accessible observables \cite{chen2021phdos,lee2026alignn,okabe2024virtual,hsu2026tsenn,wang2026deepscireasoner}, developing generalizable structure-to-spectroscopy models remains challenging, particularly in accommodating heterogeneous, high-dimensional observation domains and capturing spectral responses to complex structural variations.
More specifically, different spectroscopic observables are defined over distinct coordinate domains, ranging from one-dimensional energy-dependent spectra, such as electronic and phonon densities of states~\cite{kaundinya2022edos,lee2023dostransformer,gurunathan2023phonon,liang2024csgn,xu2026phonontransformer}, to multidimensional momentum- and energy-resolved scattering functions, such as powder and single-crystal inelastic neutron scattering (INS)~\cite{fair2022euphonic} and resonant inelastic X-ray scattering (RIXS)~\cite{degroot2024rixs}. 
Time-resolved measurements such as time-resolved RIXS and INS introduce additional temporal coordinates, further increasing the dimensionality of the observation space~\cite{dean2016ultrafast,doi:10.1107/S1600576718004727,hw9y-y2cn}. 
Conventional grid-based representations require predefined discretizations that can therefore become prohibitively expensive at higher dimensions, with the number of output values scaling as $N^d$ for a $d$-dimensional domain sampled at $N$ points per dimension.

Meanwhile, atomic structures themselves span a vast, high-dimensional configuration space, encompassing variations in chemical composition, atomic arrangements, lattice geometry, and structural disorder~\cite{musil2021representations}. Beyond idealized periodic crystals, experimentally relevant materials often exhibit defects~\cite{freysoldt2014defects}, substitutional disorder~\cite{keen2015disorder,sheriff2024motifs}, strain distributions~\cite{qi2023strain}, and complex superstructures, such as moir\'{e} patterns~\cite{kennes2021moire}, that require increasingly large atomic configurations for an explicit description~\cite{deringer2021disordered,li2022deeph,gong2023deehe3}. 
Beyond reproducing characteristic spectra, accurate structure-to-spectroscopy models must resolve how subtle structural variations shift excitation energies and redistribute spectral intensity~\cite{vizoso2023manifold,hsu2026tsenn}.
Agreement between predicted and reference spectra for individual structures does not, by itself, establish whether a model captures these structure-induced spectral changes.

Learning spectroscopic responses therefore requires navigating two complex spaces simultaneously: the atomic configuration space $\mathcal{X}$ and the spectroscopic observation space $\Omega$. The challenge is to learn a mapping $S:\mathcal{X}\times\Omega\rightarrow\mathbb{R}$ that captures how variations in atomic structure reshape spectroscopic responses across the observation domain.
Broad pretraining across materials and subsequent adaptation to specific structural families offer a promising approach~\cite{batatia2025foundation,yuan2026foundations,radova2025finetuning,grunert2025optimate}, particularly when accurate prediction requires resolving subtle spectral changes beyond characteristic spectral patterns. 
Beyond predictive accuracy, spectroscopic supervision may also provide physically meaningful representations~\cite{kong2022mat2spec,liang2023latent,yang2026matbind} that organize materials according to their observable-dependent similarities~\cite{isayev2015cartography,kuban2022dos,zadoks2024sorep}. 
A computationally scalable framework that supports both complete-spectrum prediction and targeted analysis would further help bridge idealized calculations and increasingly realistic material configurations.
These needs call for a computationally scalable framework that connects atomic configurations to spectroscopic responses across diverse observation domains, supports adaptation to structural variations, and enables both complete-spectrum prediction and targeted spectral analysis.

Here, we introduce a framework for \textbf{Spec}troscopy with \textbf{S}tructure-\textbf{Co}nditioned \textbf{Ne}ural Fields (SpecSCoNe), which learns spectroscopic responses directly from atomic structures.
By coupling an equivariant encoder~\cite{liao2026equiformerv3} that maps atomic configurations $X\in\mathcal X$ to learned material representations with a coordinate-conditioned neural network~\cite{park2019deepsdf,sitzmann2020siren,xie2022neuralfields,dupont2022functa,chitturi2023dynamical,ni2026wien} operating over the spectroscopic observation space $\Omega$, our approach learns continuous spectroscopic response functions that can be evaluated at arbitrary observation coordinates (Section~\ref{sec:field}).
We demonstrate accurate spectral prediction across different observation domains and dimensionalities using a common structure-conditioned neural-field formulation, providing a flexible alternative to conventional fixed-grid representations (Section~\ref{sec:fidelity}).
Beyond reproducing characteristic spectra, we further investigate the model's adaptability to structural variations, showing that sparse adaptation captures strain- and composition-dependent spectral responses and enables transfer to larger atomic configurations (Section~\ref{sec:response}).
The resulting material representations also provide a basis for examining how spectroscopic supervision organizes materials according to observable-dependent similarities (Section~\ref{sec:representation}).
To connect these predictive capabilities with practical spectral analysis, we demonstrate the computational scalability of direct spectral prediction and the flexibility of coordinate-based evaluation for analysis of a targeted region of the spectrum without requiring construction of the full spectrum (Section~\ref{sec:analysis}).
Our work establishes a generalizable and computationally scalable framework for learning spectroscopic response functions, with potential applications in materials characterization, spectroscopy-guided inverse problems, and experimental design.

%% file: sections/results_fig1.tex
\subsection{A structure-conditioned spectral field}\label{sec:field}

\mainfig{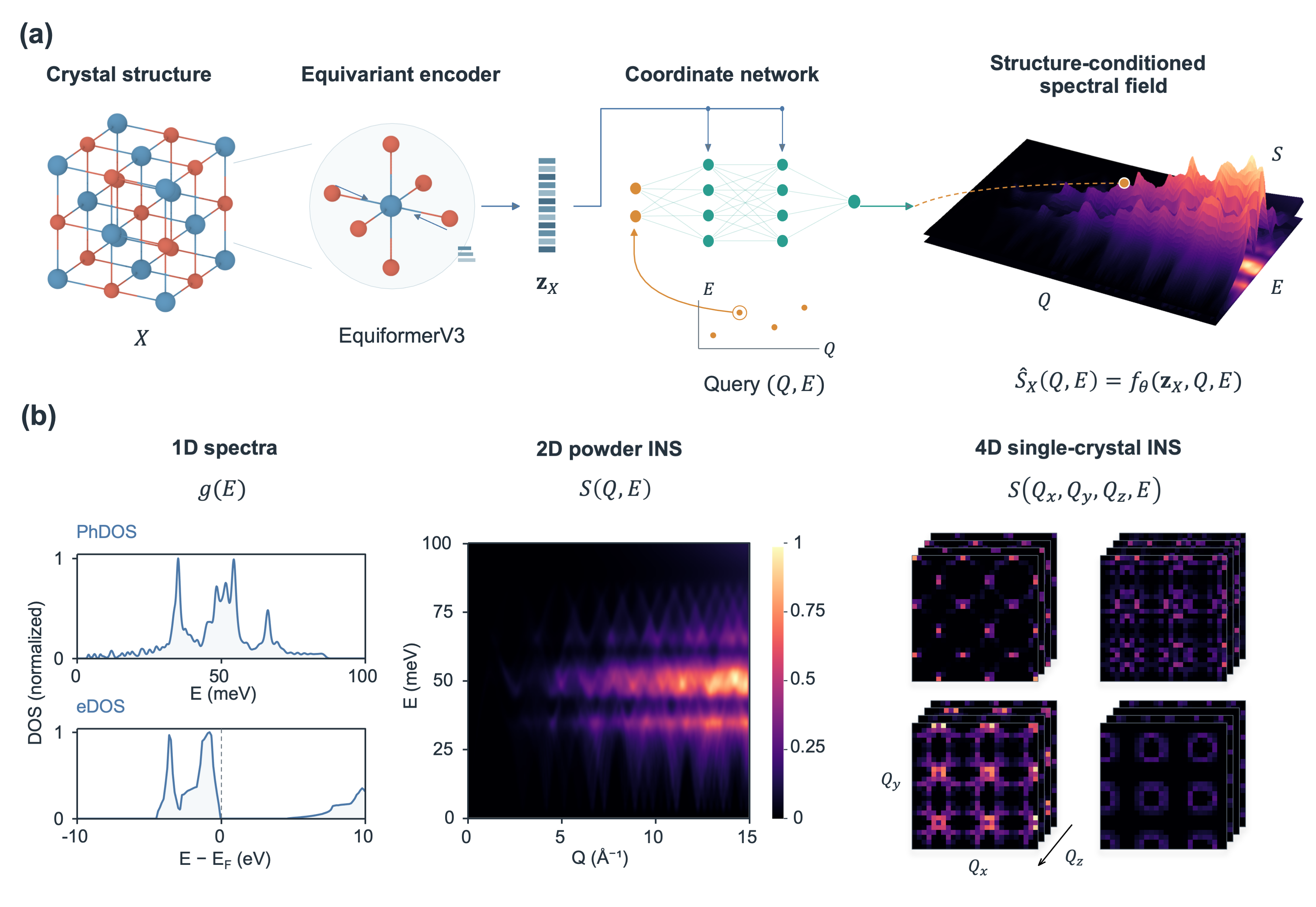}{0.65}

To learn how atomic configurations determine spectroscopic responses across distinct observation domains, we represent these responses as structure-conditioned neural fields (Fig.~\ref{fig:1}a). SpecSCoNe parameterizes each response as a function of observation coordinates, allowing a trained model to evaluate spectral intensities at requested locations rather than restricting its output to a predefined grid. For a spectroscopic observable indexed by \(m\), the predicted response is
\[
\widehat{S}_m(X,\boldsymbol{\xi}_m) = f_{\theta_m}\!\left(\mathbf z_X,\boldsymbol{\xi}_m\right),
\]
where \(X\in\mathcal X\) denotes an atomic configuration, \(\mathbf z_X\) is the encoder learned material representation, and \(\boldsymbol{\xi}_m\in\Omega_m\) specifies the observation coordinates. The coordinate network \(f_{\theta_m}\), with trainable parameters \(\theta_m\), takes the material representation and these coordinates as inputs and returns the predicted spectral value at that location. We first apply this formulation to powder inelastic neutron scattering, for which averaging over crystallite orientations gives a scattering intensity that depends on the momentum-transfer magnitude \(Q=|\mathbf Q|\) and energy transfer \(E\). In this case, we have \(\boldsymbol{\xi}_{\mathrm{INS}}=(Q,E)\) for $m=\mathrm{INS}$.

For powder INS, a pretrained equivariant graph neural network processes the periodic crystal graph, and its atomic features are aggregated into the material representation \(\mathbf z_X\). The structural encoder and coordinate network are trained jointly against reference spectra, allowing spectral supervision to refine the structural representation according to the information needed to reproduce the momentum- and energy-dependent scattering response. The framework accommodates different choices of graph-based structural encoders and coordinate networks. Here, we implement these components using EquiformerV3~\cite{liao2026equiformerv3} and a sinusoidal representation network (SIREN)~\cite{sitzmann2020siren}, respectively (Methods).

Once a crystal is encoded, its material representation can be reused to evaluate the spectral response at different observation coordinates without repeating the structural encoding. Evaluating a complete \(Q\)--\(E\) grid reconstructs a powder spectrum, whereas restricting queries to a fixed-\(Q\) line or a selected momentum--energy region provides targeted spectral information without generating the full map. 

To test whether powder INS spectrum reconstruction depends on a particular network implementation, we evaluated alternative combinations of structural encoders and spectral decoders. Across twelve combinations of four graph encoders and three decoders, the models retained high reconstruction fidelity, indicating that successful structure-to-spectrum prediction is not confined to the EquiformerV3--SIREN implementation (Supplementary Note~\ref{note:f1}).

The same structure-conditioned neural-field formulation extends naturally from powder INS to a range of one-dimensional (1D), energy-dependent observables probing distinct properties.
As shown in Fig.~\ref{fig:1}b, we demonstrate this flexibility by pairing the same graph network encoder architecture with a coordinate network trained separately for each observable: phonon density of states (PhDOS), neutron-weighted density of states (NWDOS), 1D INS (as measured by vibrational spectrometers such as VISION \cite{cheng2019oclimax}), and electronic density of states (eDOS), using energy as the observation coordinate. These targets describe vibrational or electronic state distributions, or neutron-scattering intensity under a specified measurement geometry. Although the target responses and trained parameters differ, each model retains the same separation between the learned material representation and coordinate-dependent spectral readout.

Beyond these one-dimensional observables, we examine whether coordinate networks can also represent higher-dimensional scattering responses. Single-crystal INS retains the directional dependence of momentum transfer, yielding a four-dimensional scattering response \(S(Q_x,Q_y,Q_z,E)\). The growing size of dense momentum--energy grids motivates coordinate-based representations, provided they can capture the structured intensity variations across all four dimensions. To assess this representational capacity, we fitted a Fourier--sine coordinate network~\cite{sitzmann2020siren,tancik2020fourier} to a phonon-derived single-crystal INS reference spectrum for crystalline neon (Ne), jointly optimizing the network parameters and a crystal-specific latent code without a structural encoder. The fitted field recovers ring- and arc-like intensity patterns in reciprocal-space sections summed over 0--10~meV, together with the principal low-energy features of the momentum-averaged spectrum (Supplementary Fig.~\ref{fig:si-capacity}). The recovery of reciprocal-space patterns and energy-dependent features supports the capacity of a coordinate network to represent structured four-dimensional scattering responses. Predicting such responses for unseen crystals will require learning transferable structure--spectroscopy relationships from a broader set of well-resolved single-crystal reference spectra.

%% file: sections/results_fig2.tex
\subsection{Spectroscopic fidelity across materials and observables}\label{sec:fidelity}
To assess generalization across materials, we evaluated SpecSCoNe’s powder-INS predictions against phonon-derived reference spectra for 1,001 held-out test materials. Over the complete \(Q\)--\(E\) grid, the predictions achieved a median per-material Pearson correlation coefficient (PCC) of 0.979 and a median normalized \(L_2\) error (N-L2) of 0.186. The learned field reproduced characteristic momentum- and energy-dependent scattering patterns across most of the test set, with 890 materials (88.9\%) reaching a PCC of at least 0.90. Validation and test error distributions were similar, and aggregate test performance varied little across three independent training runs (Supplementary Fig.~\ref{fig:si-fidelity-features}a,b; Supplementary Table~\ref{tab:si-f2-seeds}).
\mainfig{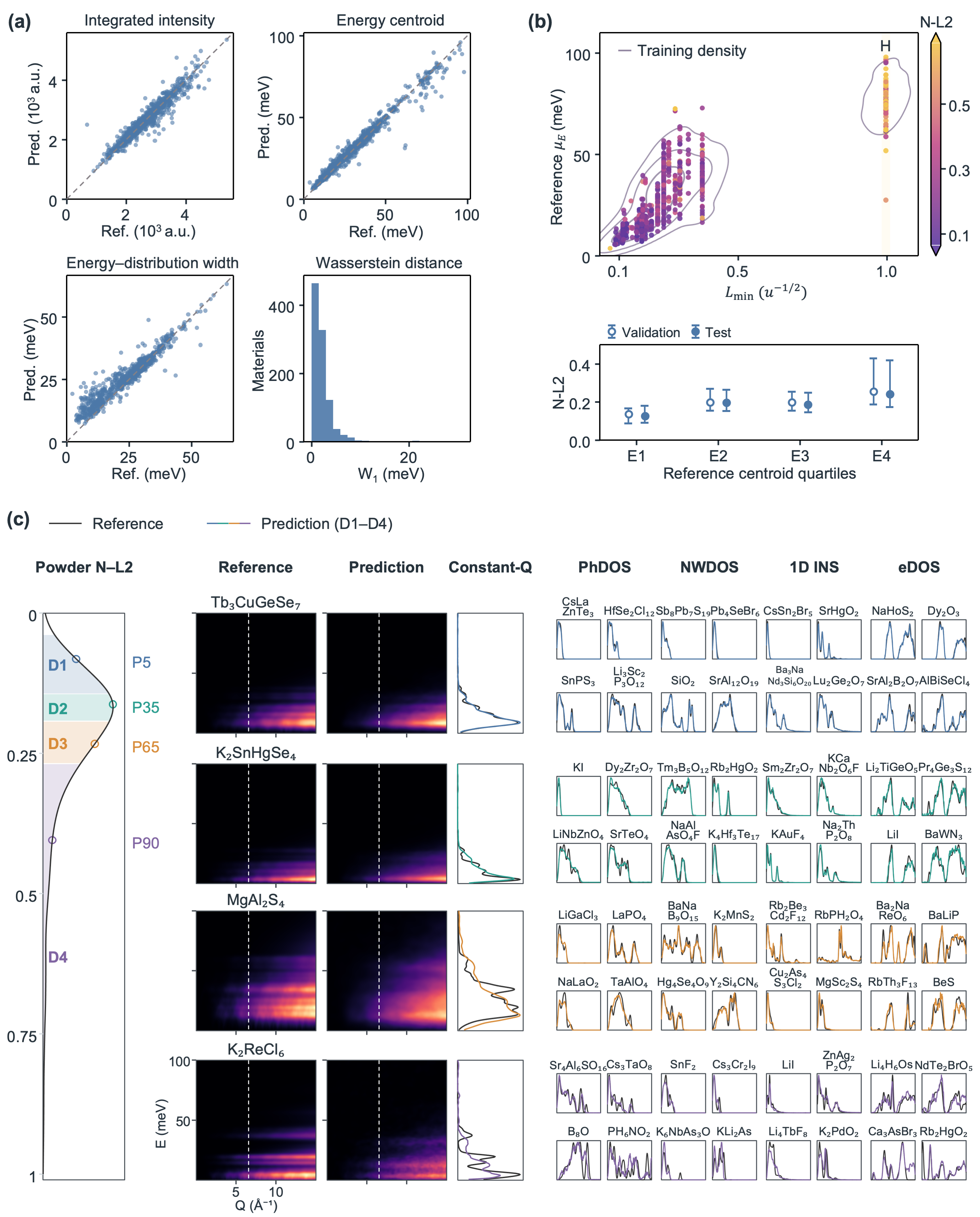}{0.70}

We next examined whether the flexibility of coordinate-based prediction comes at the expense of full-spectrum accuracy. Under matched training conditions and with comparable decoder parameter counts, SpecSCoNe achieved a mean N-L2 of 0.230 across 1,001 validation materials, compared with 0.264 for a low-rank fixed-raster decoder, with lower errors for 69.0\% of materials (Supplementary Fig.~\ref{fig:si-field-raster}). The coordinate readout thus enables selective spectral evaluation without compromising reconstruction accuracy on the full reference grid.
 
Beyond reproducing the overall \(Q\)--\(E\) scattering patterns, we examined whether the predicted spectra preserve the integrated scattering weight and its distribution over energy (Fig.~\ref{fig:2}a). In the \(Q\)-integrated profiles, the scattering weight within the evaluated domain, quantified by the integrated intensity~\cite{berk1993neutron}, was recovered with a mean absolute relative error of 4.86\% on the common reference-data intensity scale. The intensity-weighted energy centroid and energy-distribution width, which characterize the mean energy and spread of this weight, were recovered with mean absolute errors of 1.77 and 2.31~meV, respectively. Because profiles with similar centroids and widths can nevertheless differ in spectral shape, we further compared their area-normalized energy distributions using the Wasserstein-1 distance, which measures the minimum average displacement along the energy axis required to redistribute the normalized predicted scattering weight to match the reference distribution. SpecSCoNe closely reproduces the normalized energy distributions, with a median Wasserstein-1 distance of 1.60~meV across the held-out test materials.

Across the validation set, the N-L2 distribution is concentrated at low reconstruction errors, with a tail extending to larger errors (Fig.~\ref{fig:2}c). Among the examples spanning four error strata (D1--D4), the low-error predictions for Tb$_3$CuGeSe$_7$ and K$_2$SnHgSe$_4$ reproduce the positions and energy ranges of the dominant spectral bands, as well as their intensity variation with \(Q\). The corresponding constant-\(Q\) profiles also follow the principal features of the reference spectra. Even for K$_2$ReCl$_6$, which lies in the higher-error region, the predicted scattering weight remains concentrated in the correct low-energy range, although distinct reference maxima are not fully resolved. Across the validation-set \(Q\)-integrated profiles, matched peaks have a mean absolute energy error of 1.12~meV, whereas a peak recall of 0.433 reflects the greater difficulty of recovering spectral peaks under the specified detection and matching criteria (Supplementary Fig.~\ref{fig:si-fidelity-features}c; Supplementary Note~\ref{note:f2}).

Materials with higher reference energy centroids and greater compositional lightness tend to exhibit larger reconstruction errors (Fig.~\ref{fig:2}b). Grouping test materials by reference energy centroid using quartiles defined from the training distribution, we find a median N-L2 of 0.125 in the lowest-centroid group (E1), compared with 0.240 in the highest-centroid group (E4). Because vibrational frequencies depend on both atomic masses and interatomic restoring forces, with lighter masses generally supporting higher frequencies at comparable stiffness, we also examined compositional lightness, \(L_{\min}=m_{\min}^{-1/2}\), where \(m_{\min}\) is the mass of the lightest constituent element. The INS energy centroid additionally reflects how these vibrations contribute to neutron scattering, since their spectral weights depend on atomic displacement patterns and neutron-scattering strengths~\cite{fair2022euphonic}. Test N-L2 is positively correlated with \(L_{\min}\) (Spearman \(\rho=0.448\)), and this association persists when using atomic-fraction-weighted lightness or excluding hydrogen-containing materials (Supplementary Fig.~\ref{fig:si-material-controls}c).

More frequent occurrence of an element in training, however, does not consistently correspond to lower reconstruction error. Across 73 elements, the Spearman correlation between the number of training structures containing an element and the mean test N-L2 for materials containing that element was \(\rho=-0.167\), compared with \(+0.186\) when the analysis was restricted to elements represented in at least ten test materials (Supplementary Fig.~\ref{fig:si-material-controls}b). Hydrogen appears in 621 training structures, while H-containing test materials exhibit higher reconstruction errors and a reference energy-centroid distribution shifted toward higher energies than that of H-free materials (Supplementary Fig.~\ref{fig:si-hconditional}; Supplementary Note~\ref{note:f2}). Elemental occurrence alone does not capture the diversity of bonding environments and associated vibrational responses, making it an incomplete measure of the training coverage relevant to spectral prediction.

To examine prediction fidelity across the 1D observables introduced in Fig.~\ref{fig:1}b, we evaluated the separately trained PhDOS, NWDOS, 1D INS and eDOS models against Gaussian-smoothed reference spectra. Median validation N-L2 values were 0.186, 0.172, 0.163 and 0.241, respectively (Supplementary Table~\ref{tab:si-1d-validation}). Representative comparisons spanning four task-specific error strata (D1--D4) show how reconstruction quality varies across materials: predictions generally reproduce the prominent spectral features and overall envelopes, while discrepancies in finer features become more apparent in higher-error cases (Fig.~\ref{fig:2}c). Additional comparisons across each task's error distribution are provided in Supplementary Fig.~\ref{fig:si-1d-atlas}.
These results establish spectral fidelity across diverse materials and observables, raising the further question of whether the models also capture spectral changes induced by structural variations.

%% file: sections/results_fig3.tex
\subsection{Learning spectroscopic responses across structural variations and scales}
\label{sec:response}

High fidelity in spectral reconstruction across materials does not necessarily translate into accurate prediction of the subtle spectral changes induced by structural perturbations. 
A model may reproduce a material's characteristic spectrum while missing the smaller shifts and intensity changes that reveal its response to structural variation. Capturing these changes is essential for connecting spectral observations to changes in atomic structure and provides a more demanding test of the learned structure–spectroscopy relationship.

We first examined this sensitivity using biaxial strain to vary the lattice geometry at fixed composition in four bulk 2H transition-metal dichalcogenides (TMDs), MoSe$_2$, MoS$_2$, WSe$_2$ and WS$_2$, with DFT-derived spectra providing the reference responses. 
In WSe$_2$, tensile strain shifts prominent bands toward lower energies while preserving much of the overall spectral pattern (Fig.~\ref{fig:3}a). To examine these changes directly, we evaluated the finite-strain difference spectrum:
\[
\Delta S_\varepsilon(Q,E)
=S(Q,E;X_\varepsilon)-S(Q,E;X_0),
\]
where \(X_\varepsilon\) and \(X_0\) denote the strained and unstrained structures, respectively. Band shifts appear in these maps as adjacent regions of intensity gain and loss, revealing a strain-induced redistribution of scattering intensity that is less apparent in the absolute spectra.

In WSe$_2$, the difference spectrum predicted by the globally trained SpecSCoNe model blurs the alternating bands resolved in the DFT reference. Across the eight withheld configurations at $\pm0.5\%$ strain, mean response PCC was 0.621, even though the model reproduced the compounds' characteristic spectra with a mean PCC of 0.983. Fine-tuning the global model with sparse supervision from 20 training structures spanning the four compounds improved its reconstruction of these response features (Fig.~\ref{fig:3}a). After adaptation, mean response PCC increased to 0.971, while mean full-grid response N-L2 decreased from 0.827 to 0.238.
\mainfig{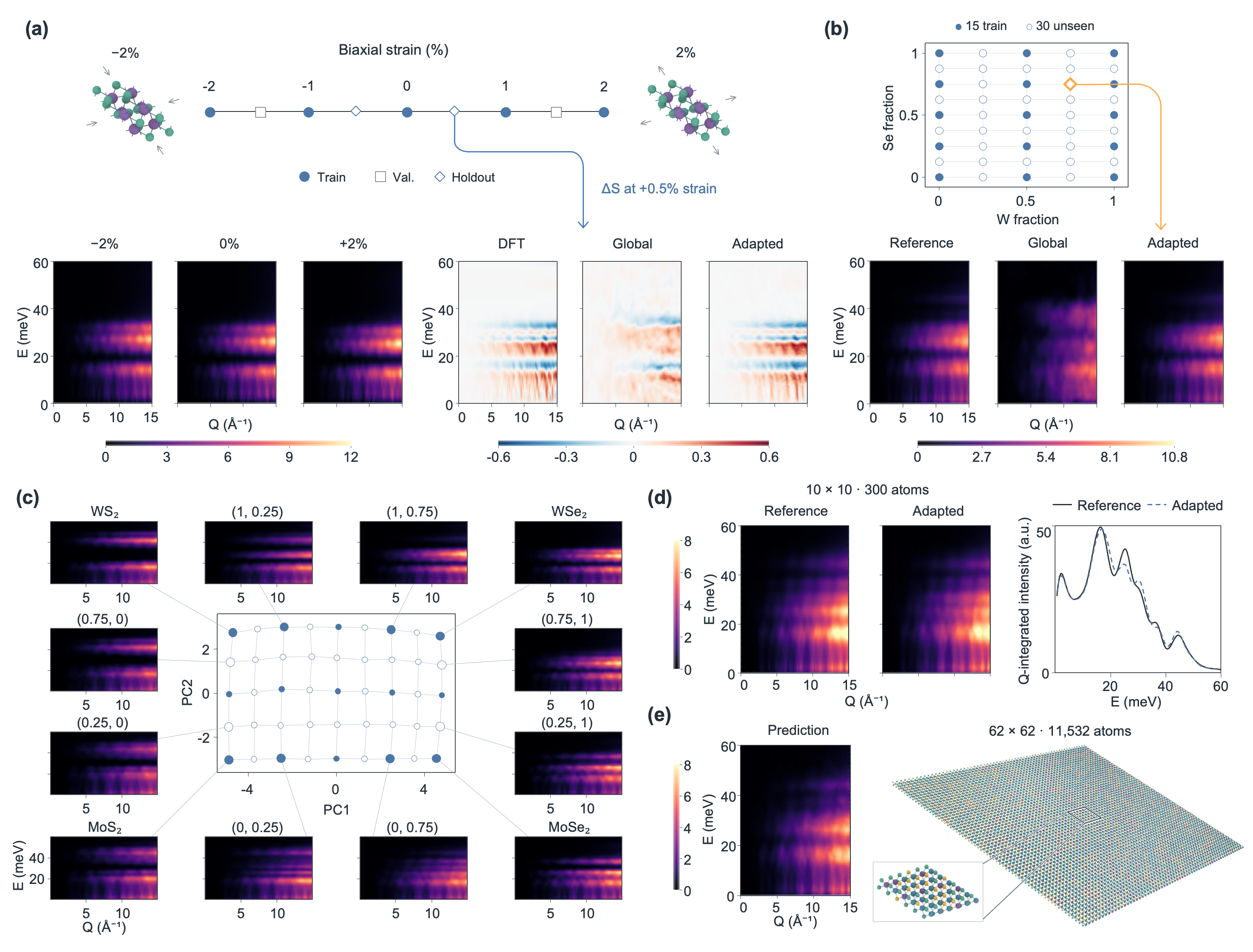}{0.66}

While strain modifies lattice geometry at fixed composition, chemical substitution also changes atomic masses, local bonding and scattering cross-sections, introducing a distinct source of spectral variation. We examined this dependence in 12-atom monolayer configurations of $\mathrm{Mo}_{1-x}\mathrm{W}_x\mathrm{S}_{2(1-y)}\mathrm{Se}_{2y}$, using reference spectra derived from a specialized MACE interatomic potential~\cite{siddiqui2024alloys} (Methods). 
We note here that these monolayer configurations serve as a computational test of composition-dependent spectral prediction, although direct INS measurements on isolated monolayers are not experimentally feasible.
A separate model initialized from the global model was adapted on 15 compositions and evaluated on the remaining 30. At the withheld composition $(x,y)=(0.75,0.75)$, the global model places prominent spectral weight too high in energy and blurs the separation between bands, whereas the adapted prediction more closely reproduces the reference band positions and relative intensities (Fig.~\ref{fig:3}b). The improved spectral reconstruction extends across the 30 withheld compositions, with median full-grid N-L2 decreasing from 0.411 to 0.0924 and median PCC increasing from 0.912 to 0.995 (Supplementary Note~\ref{note:f3}).

Compositional variation is also reflected in the material representations learned by the composition-adapted model, with each representation locating an atomic configuration in a learned materials space. In a two-dimensional principal component analysis (PCA) projection capturing 97.5\% of the variance across the sampled material representations, both training and withheld compositions follow systematic trends between the four endmembers along the Mo--W and S--Se substitution series (Fig.~\ref{fig:3}c). The spectra predicted from these representations vary systematically along these compositional series: increasing Se content along both the Mo- and W-based boundaries shifts prominent high-energy scattering weight toward lower energies, while intermediate compositions redistribute intensity among several bands. Within this alloy family, compositional variation is therefore reflected both in the organization of the learned materials space and in the evolution of the predicted scattering responses (Supplementary Note~\ref{note:f3}).

Whereas the preceding analyses examine strain and chemical substitution in small periodic cells, explicitly describing defects, substitutional disorder and long-period superstructures can require atomic configurations spanning many unit cells~\cite{freysoldt2014defects,sheriff2024motifs,li2022deeph}. Predicting their spectroscopic responses therefore calls for models that retain predictive accuracy beyond the small configurations used for adaptation. We examined this size-transfer capability within the substitutional-alloy family, applying the composition-adapted SpecSCoNe model without further training to independently randomized and relaxed configurations containing up to 300 atoms. At the withheld composition $(x,y)=(0.25,0.625)$, the prediction for the $10\times10$, 300-atom configuration reproduces the dominant spectral bands and the major peaks and shoulders of the $Q$-integrated profile, with a full-grid N-L2 of 0.0725 and PCC of 0.9968 (Fig.~\ref{fig:3}d). Across two compositions and five configuration sizes, full-grid N-L2 remains between 0.0595 and 0.0938 (Supplementary Table~\ref{tab:si-size-transfer}), with spectral fidelity maintained beyond the small periodic configurations used for adaptation.


The agreement with reference spectra across independently generated configurations containing up to 300 atoms provides evidence that the composition-adapted model transfers beyond the small cells used for training. At larger scales, however, validating this transfer becomes increasingly difficult because conventional force-based workflows require expensive force-constant construction followed by phonon and neutron-scattering calculations~\cite{han2024inspired}. To explore the computational reach of direct spectral prediction, we applied the same model without further adaptation to a $62\times62$ substitutional-alloy configuration containing 11,532 atoms, generating a complete powder-INS spectrum in just 2.83~s on a single GPU (Fig.~\ref{fig:3}e; see Section~\ref{sec:analysis} for timing details). Generating a reference spectrum for this configuration was prohibitively expensive under the reference workflow, so its spectral accuracy remains unverified (Methods; Supplementary Note~\ref{note:f3}). Nevertheless, the validated transfer through 300 atoms provides a basis for exploring these larger configurations. By bypassing the intermediate calculations that limit conventional workflows, SpecSCoNe makes spectral prediction computationally accessible for disordered configurations containing on the order of $10^{4}$ atoms.

%% file: sections/results_fig4.tex
\subsection{Spectroscopic supervision reorganizes material neighborhoods}
\label{sec:representation}
The composition-adapted representations in Fig.~\ref{fig:3}c organize compositional variations within an alloy family along well-defined substitution directions. 
More generally, by mapping atomic structures of different sizes and compositions into a fixed-dimensional space, the encoder provides a common basis for comparing materials.
For chemically and structurally diverse compounds beyond a single substitution series, however, the physical significance of proximity in a learned materials space requires further examination. 
Although compositional and crystallographic resemblance provide natural criteria for comparing materials, compounds differing in both can exhibit similar electronic densities of states~\cite{kuban2022dos}. 
Such spectroscopic similarity depends on the observable being compared: here, powder INS characterizes the neutron-scattering response of lattice vibrations, whereas eDOS describes the energy distribution of electronic states. In the reference spectra shown in Fig.~\ref{fig:4}a, BaCl$_2$ resembles PrIn$_2$Cl$_5$ in powder INS, whereas its eDOS more closely resembles that of LiI. We therefore examined whether learning to predict different spectroscopic observables reorganizes structure-derived material representations according to similarity in the corresponding spectra.
\mainfig{4}{0.66}

For BaCl$_2$, the choice of spectroscopic supervision changes how other compounds are ranked by cosine distance in the learned representation space (Fig.~\ref{fig:4}b). Under matched training conditions (Methods), INS supervision raises PrIn$_2$Cl$_5$ from rank 77 at initialization to rank 3 among BaCl$_2$'s candidate neighbors, placing the compound with a similar powder spectrum among its closest neighbors. Under eDOS supervision, LiI rises from rank 536 to rank 61, consistent with its electronic-spectrum similarity to BaCl$_2$. Although the models are trained to predict spectra without an explicit objective for pairwise material similarity, these spectrally similar compounds become more highly ranked under the corresponding supervision.

To determine whether this behavior extends across materials, we retrieved the five nearest neighbors of each of 649 held-out test materials by representation-space cosine distance and evaluated their similarity to the query using reference spectra that were not involved in neighbor selection. We quantified this similarity by neighborhood enrichment, defined as the fractional reduction in mean reference-spectral distance relative to randomly selected neighbors (Methods). INS supervision increases INS enrichment from 0.361 at initialization to 0.445, while eDOS supervision increases eDOS enrichment from 0.157 to 0.193 (Fig.~\ref{fig:4}c). The improvement also extends to identifying specific spectrally similar compounds, as matching supervision increases the overlap between the neighbor sets defined by the learned representations and those defined by the corresponding reference spectra (Supplementary Note~\ref{note:f4}).

Evaluating these same representation-selected neighborhoods against the alternative observable reveals the selectivity of the improvement. In the five-neighbor comparison, INS supervision slightly reduces eDOS enrichment, whereas eDOS supervision leaves INS enrichment close to initialization (Fig.~\ref{fig:4}c). This preference is also reflected in which compounds enter or leave the learned neighborhoods. Among queries whose neighbor sets change, newly retrieved neighbors under INS supervision have, on average, closer INS spectra but more distant eDOS spectra than the materials they replace. Under eDOS supervision, the replacements have closer eDOS spectra without a clearly resolved change in INS similarity (Supplementary Note~\ref{note:f4}).

Joint INS--eDOS supervision allows one material representation to support neighborhoods with enhanced similarity in both spectra. Across the test materials, evaluating the same five-neighbor sets against both reference spectra gives INS and eDOS enrichment values of 0.387 and 0.191, respectively, both above initialization (Fig.~\ref{fig:4}c). Relative to single-task training, the joint model nearly matches the eDOS-only enrichment of 0.193 but gives lower INS enrichment than the INS-only value of 0.445 (Supplementary Note~\ref{note:f4}). A shared representation can therefore preserve both spectral relationships, although the present joint training does not simultaneously attain the enrichment achieved by the two separately supervised models.
These results show that a shared, fixed-dimensional representation can capture material similarities across distinct spectroscopic observables, providing a basis for spectroscopy- and property-guided materials discovery and a potential starting point for inverse materials design.

\subsection{Querying representations for spectroscopic analysis}
\label{sec:analysis}
Quantitative analysis of vibrational spectra often focuses on characteristic energies and the distribution of scattering intensity within selected spectral regions. In single-crystal INS, energy profiles extracted from selected momentum regions have been used to analyze phonon dispersions and intensity exchange between coupled phonon branches~\cite{garlatti2020phonons}. Spatially resolved vibrational electron spectroscopy similarly characterizes defect-induced changes through vibrational energy shifts and intensities integrated over selected energy windows~\cite{yan2021singledefect}. For quantities defined within selected spectral regions, SpecSCoNe enables the sampling domain and density to be tailored to the analysis without first generating a complete spectral grid.

We first used targeted coordinate queries to characterize local scattering-weight distributions in the alloy series of Fig.~\ref{fig:3}, where chemical substitution changes the positions and relative intensities of vibrational spectral features. With the composition-adapted model, we evaluated intensities at five \(Q\) values and eighteen energies within \(Q=4\)--8~\AA$^{-1}$ and \(E\simeq10\)--60~meV, requiring only 90 coordinate queries per structure (Fig.~\ref{fig:4}d). For the illustrated withheld composition, the resulting local \(Q\)-integrated energy profile follows the main features of the reference profile. From these intensities, we extracted the local energy centroid to compare the intensity-weighted mean energy of the response across compositions. We also calculated the fraction of the window-integrated scattering weight above 30~meV to quantify the relative contribution of the selected higher-energy region.

Tracking a spectral feature across perturbed structures provides a quantitative measure of its sensitivity to lattice geometry. For the biaxially strained TMD series introduced in Fig.~\ref{fig:3}, we quantified this sensitivity by tracking selected powder-intensity ridges and extracting a finite-difference coefficient \(\alpha\), which measures the ridge-energy shift per percentage point of biaxial strain. The reference ridges shift toward lower energies under tension in all four compounds, but their response magnitudes differ, with coefficients ranging from \(-0.387\)~meV/\% in WS$_2$ to \(-0.687\)~meV/\% in MoS$_2$. Using a strain-adapted model (Methods), we extracted the corresponding coefficients from predicted spectra at the withheld \(\pm0.5\%\) strain states. The predicted coefficients recover both the shift direction and the ordering of the response magnitudes, with a mean absolute error of 0.044~meV/\% across the four compounds (Fig.~\ref{fig:4}e; Supplementary Note~\ref{note:f4}).

Energy shifts alone do not quantify how strain redistributes scattering intensity around a spectral feature. Within a selected momentum--energy window, intensity gains and losses can cancel on integration, masking changes in the local spectral profile. We therefore quantified the scattering weight gained and lost separately by integrating positive intensity changes and the magnitudes of negative changes relative to the unstrained spectrum, using fixed windows around the tracked ridges. Across the eight withheld strain states, the predicted gains and losses correlate with their DFT-derived references with Pearson coefficients of 0.984 and 0.993, respectively (Supplementary Note~\ref{note:f4}).

Quantitative spectral analysis remained accurate even when only a small fraction of the full spectral grid was sampled (Fig.~\ref{fig:4}f). Across the 30 withheld alloy compositions, just 90 coordinate queries per structure, approximately 0.1\% of the full grid, yielded local energy centroids and high-energy intensity fractions with mean absolute errors of 0.108~meV and 1.07 percentage points, respectively, against the reference values (Supplementary Table~\ref{tab:si-query-budgets}). Sparse sampling also retained close agreement with the densely sampled strain-response coefficients, with a mean absolute deviation of 0.00235~meV/\% when the stored validation spectra were subsampled to approximately 0.19\% of the grid points per state (Supplementary Note~\ref{note:f4}). Although denser sampling can further reduce numerical deviations from the full-grid estimates, the small errors already reached at these budgets allow the sampling density to be chosen for the precision required by the local analysis, without requiring dense evaluation of the entire spectrum.

SpecSCoNe also provides an efficient route to predicting complete powder-INS maps directly from atomic structures. In conventional force-based workflows, forces evaluated for displaced structures are used to construct interatomic force constants, from which phonon modes and neutron-scattering intensities are calculated. For the same already-relaxed WSe$_2$ structure, the DFT-based route required approximately 4.34~h in summed force-job time and subsequent spectral post-processing (Fig.~\ref{fig:4}g, top; Methods). Replacing DFT with an EquiformerV3 machine-learning force field (MLFF) reduced the time for force evaluation and force-constant construction to 0.366~s, but the remaining spectrum-generation steps still required approximately 9.1~s, giving a total of 9.42~s. By bypassing force-constant construction and the subsequent phonon and neutron-scattering calculations, our model generated the complete powder-INS map in 21.7~ms, including graph construction, structural encoding and full-grid evaluation.

This computational advantage is particularly relevant to substitutionally disordered alloys, where larger supercells are needed to represent distinct atomic arrangements. Across the 12--300-atom alloy configurations, scattering post-processing dominated the recorded MLFF-based reference workflow with coherent scattering included, whereas complete-spectrum prediction by our model required approximately 32--84~ms (Fig.~\ref{fig:4}g, bottom). The cost of reference generation, however, depends on the numerical settings and scattering approximation. Additional calculations for a 300-atom configuration showed that reducing the phonon mesh preserved the principal spectral bands at lower cost, while the faster incoherent approximation introduced larger differences in local momentum-dependent features (Supplementary Fig.~\ref{fig:si-oclimax-protocol}). 
Direct prediction remained computationally practical at larger scales: predicting the complete powder-INS map for the 11,532-atom configuration discussed in Section~\ref{sec:response} required only 2.83~s. This inference speed could enable on-the-fly comparison of predicted spectra with experimental measurements, helping assess candidate atomic configurations and guide subsequent measurements in real time.

%% file: sections/discussion.tex
\section{Discussion}
We presented a generalizable and computationally scalable framework for learning structure-dependent spectroscopic responses across diverse atomic configurations and observation domains. Beyond accurate spectral prediction, the framework offers a route toward connecting atomistic descriptions with spectroscopic observations of increasingly complex materials and spectra. We demonstrated that broad pretraining followed by sparse adaptation enables the model to capture subtle spectral changes induced by strain and chemical substitution, while extending predictions from small training structures to larger configurations with explicit compositional disorder. By bypassing computationally expensive intermediate calculations, direct spectral prediction makes it more feasible to investigate how local atomic environments and structural variations manifest in observable spectra. Meanwhile, coordinate-conditioned evaluation allows spectral quantities to be extracted from selected regions of the observation space, a particularly useful capability for experimental spectroscopy, where accessible coordinates are often constrained by scattering geometry, detector coverage, or missing measurements. Beyond forward prediction, spectroscopic supervision also shapes the learned material representations, providing a basis for identifying materials with similar physical responses according to the observable of interest, rather than relying solely on structural or compositional similarity. These capabilities suggest opportunities for connecting atomistic modeling more closely with experimental characterization and materials discovery.

Despite these advances, several challenges remain. First, while the framework accommodates spectroscopic observables with different dimensionalities, generalization across materials for high-dimensional single-crystal spectra remains limited by the availability and resolution of training data. Our four-dimensional demonstration establishes the capacity of coordinate-based neural networks to represent complex momentum- and energy-dependent scattering functions \cite{liu2026observation}. However, learning their dependence on atomic structure requires sufficiently diverse and well-resolved reference spectra, which are substantially more expensive to generate than lower-dimensional observables. The limited coverage and relatively coarse resolution of the available single-crystal training data may therefore constrain the model's ability to learn transferable structure–spectroscopy relationships. Addressing this challenge will require broader high-dimensional spectroscopic datasets and potentially more effective approaches to structural representation learning.
Second, our current demonstrations focus primarily on vibrational spectroscopy, with electronic density of states providing an additional observable for representation learning. Extending the framework to a broader range of spectroscopic modalities, including electronic, magnetic, and time-resolved responses, will require more diverse reference data and potentially modality-specific physical constraints. Finally, the present models are trained predominantly on calculated spectra. Applying them to experimental measurements will require accounting for instrumental resolution, background signals, measurement noise, and incomplete observation coverage.

These limitations also point toward several opportunities for extending the framework. First, combining structure-conditioned neural fields with spectral compression techniques \cite{ni2026wien} could provide a more efficient approach to learning high-dimensional spectroscopic responses, particularly when reference data are limited or sparsely sampled. Expanding the framework to additional spectroscopic modalities and incorporating experimental measurements would further improve its applicability to realistic characterization tasks. The continuous coordinate-based formulation provides a natural interface for uncertainty-aware experimental design, where informative observation coordinates can be selected to distinguish competing material structures or physical models \cite{liu2026observation}. Finally, extending the framework from simulated to experimental spectroscopic data presents an important opportunity. By leveraging broad pretraining on computational spectra and subsequent fine-tuning with limited experimental measurements, the framework could provide adaptable spectroscopic models that account for instrument-specific responses and experimental conditions. Such capabilities would help bridge computational predictions and experimental characterization, thereby enabling more reliable interpretation of measured spectra.

The broader significance of this framework lies in establishing a more flexible and scalable connection between atomic configurations and spectroscopic observations, accommodating increasingly complex material structures and diverse experimental observation domains. Rather than being constrained by predefined spectral grids, the observation coordinates can be determined by the scientific question or experimental conditions. This flexibility is particularly relevant to multidimensional scattering experiments, where measurements often cover only selected regions or trajectories in momentum--energy space. By enabling rapid evaluation of spectroscopic responses at experimentally accessible coordinates, the framework opens opportunities for direct comparison between measured signals and atomistic models that explicitly account for strain, disorder, and other structural variations, without requiring complete spectral acquisition. 
Combined with its computational efficiency, this capability could allow candidate atomic configurations to be assessed on-the-fly against incoming measurements during an experiment, helping identify plausible atomistic interpretations and guide the selection of measurement regions that yield informative data.
Such integration could move structure-to-spectroscopy learning beyond offline prediction toward a closed-loop framework that unifies ML-guided experimentation and experiment-informed modeling, ultimately accelerating materials characterization and discovery.

%% file: methods_revision.tex
\section{Methods}

\subsection{Spectroscopic datasets and reference calculations}\label{sm:data}
The vibrational dataset comprised 10,023 inorganic crystal structures and their calculated spectra from the published database~\cite{cheng2023insdatabase}. Reference powder spectra were generated from DFT-derived force constants from phonondb@kyoto-u~\cite{togo_phonondb} using Phonopy to obtain phonon frequencies and eigenvectors, followed by OCLIMAX neutron-scattering calculations~\cite{togo2023phonopy,cheng2019oclimax}. We used the zero-temperature powder spectra calculated over $Q=0$--15~\AA$^{-1}$ and $E=0$--150~meV with an energy-resolution setting of 1.5~meV, stored as arrays of 300 energy bins and 299 momentum-transfer bins. Structures were partitioned by recorded chemical formula into 8,021 training, 1,001 validation and 1,001 test materials, with each formula group confined to a single partition.

The PhDOS, NWDOS and 1D INS profiles contained 2,999 energy points spanning approximately 0.1--299.9~meV, with the 1D INS targets taken from the zero-temperature backward-scattering channel. The single-structure fitting experiment used a crystalline Ne reference from a separate PhononDB-derived single-crystal INS collection (Supplementary Note~\ref{sm:4d-capacity}).

Electronic spectra were obtained from 6,773 Materials Project records~\cite{jain2013materialsproject}. The spin-summed eDOS profiles were expressed per atom, aligned to the Fermi energy and linearly resampled onto 2,001 points over $[-10,10]$~eV. Standalone eDOS prediction used a formula-grouped partition of 5,421 training, 676 validation and 676 test materials.

One-dimensional training targets were Gaussian-smoothed using full widths at half maximum of 8~meV for PhDOS, NWDOS and 1D INS, and 0.8~eV for eDOS. Each smoothed profile was rescaled to preserve its original integral over the complete energy grid.

Biaxially strained configurations were generated from six-atom bulk 2H cells of MoS$_2$, MoSe$_2$, WS$_2$ and WSe$_2$ at nine equally spaced strain values spanning $-2\%$ to $+2\%$. Both in-plane lattice vectors were scaled by $1+\varepsilon$, and internal atomic coordinates were relaxed with all lattice vectors held fixed. DFT calculations used PBE with D3 dispersion corrections~\cite{perdew1996pbe,grimme2010d3}. Finite-displacement forces were used to construct interatomic force constants and calculate phonon frequencies and eigenvectors with Phonopy, followed by neutron-scattering calculations with OCLIMAX. Native-grid spectra were used for the full-spectrum response comparison, whereas the ridge-shift and gain/loss analyses used separately broadened and resampled targets derived from the same DFT force calculations (Supplementary Note~\ref{note:f3}).

Composition-dependent reference spectra were generated for the monolayer alloy series $\mathrm{Mo}_{1-x}\mathrm{W}_x\mathrm{S}_{2(1-y)}\mathrm{Se}_{2y}$, using one 12-atom configuration at each of 45 compositions spanning $0\leq x,y\leq1$, with increments of $\Delta x=0.25$ and $\Delta y=0.125$. Structural relaxation at fixed lattice vectors and finite-displacement force calculations used a MACE interatomic potential developed for Mo--W--S--Se alloys~\cite{batatia2022mace,siddiqui2024alloys}. Reference spectra were then obtained using the Phonopy--OCLIMAX workflow described above, with numerical settings specified in Supplementary Note~\ref{note:f3}.

Size-transfer references were generated for alloy configurations at $(x,y)=(0.5,0.5)$ and $(0.25,0.625)$, corresponding to a training composition and a withheld composition, respectively. For each composition and lateral size $n\times n$, with $n=2,4,6,8,10$, one independently randomized configuration was relaxed, giving ten configurations containing 12--300 atoms. Reference spectra were calculated using the same MACE--Phonopy--OCLIMAX workflow. An additional, independently randomized $62\times62$ configuration containing 11,532 atoms at $(x,y)=(0.25,0.625)$ was used for inference benchmarking.

Native-grid reference spectra for strain, composition and size transfer were calculated at zero temperature using energy and momentum resolution parameters of 1.5~meV and 0.05~\AA$^{-1}$, respectively. Intensities were normalized by the summed atomic neutron-scattering cross sections in each cell. Supplementary Table~\ref{tab:si-protocol-map} summarizes the datasets, model instances and observation domains.

\subsection{SpecSCoNe architecture}\label{sm:encoder}
For powder INS prediction, we implemented SpecSCoNe using an EquiformerV3 equivariant graph-attention encoder and a SIREN coordinate network~\cite{sitzmann2020siren}. The encoder was initialized from the gradient-force variant of EquiformerV3, pretrained on OMat24 and fine-tuned on MPtrj and subsampled Alexandria~\cite{liao2026equiformerv3}. Each crystal was represented as a periodic atomic graph with a radial cutoff of 6~\AA\ and a maximum of 300 neighbors per atom. Atomic features were extracted from the final four equivariant transformer blocks. Rotationally invariant descriptors were constructed by retaining the scalar ($L=0$) channels and taking the root-mean-square over the $2L+1$ components within each higher-degree ($L>0$) feature channel.

The blockwise descriptors were concatenated into 2,560-dimensional atomic feature vectors and projected to 256 dimensions using PCA fitted once to training-set atomic features from the pretrained encoder, before spectroscopic fine-tuning. The projected features were standardized component-wise using training-set means and standard deviations, with both transformations held fixed during subsequent training. Four-head, permutation-invariant attention pooling aggregated the standardized atomic features. Each head formed a weighted mean using weights obtained from a learned linear scoring function followed by a softmax over atoms. The concatenated head outputs were mapped to the 256-dimensional material representation $\mathbf z_X$ by a post-pooling projection multilayer perceptron (MLP), with dimensions $1{,}024\rightarrow512\rightarrow512\rightarrow256$, sigmoid linear unit (SiLU) hidden activations and output layer normalization~\cite{ba2016layernorm}.

The coordinate network was conditioned on $\mathbf z_X$ through feature-wise linear modulation (FiLM)~\cite{perez2018film}. Query coordinates $Q$ and $E$ were independently rescaled to $[-1,1]$ using fixed training-domain bounds for both full-grid and local evaluation. The SIREN comprised five width-512 sinusoidal layers---one input layer and four hidden layers---with material-dependent modulation applied before each activation:
\begin{equation}
 h^{\ell+1}=\sin\!\left\{\omega_0\left[(1+\gamma_\ell(\mathbf z_X))\odot
 (W_\ell h^\ell+b_\ell)+\beta_\ell(\mathbf z_X)\right]\right\}.
\end{equation}
Here $h^0$ contains the normalized query coordinates, $W_\ell$ and $b_\ell$ are trainable weights and biases, $\odot$ denotes element-wise multiplication, and $\omega_0=30$. A conditioning MLP with three 512-unit hidden layers with rectified linear unit (ReLU) activations generated the layer-specific modulation vectors $\gamma_\ell$ and $\beta_\ell$ from $\mathbf z_X$, which were shared across all query coordinates for a given crystal. A linear output layer predicted standardized intensity, which was transformed back to the reference scale using the global training-set intensity mean and standard deviation.

\phantomsection\label{sm:1d}\label{sec:si-1d}
For standalone one-dimensional models, energy was the sole observation coordinate and was rescaled to $[-1,1]$ over each task's fixed domain. Spectra were parameterized as nonnegative, unit-area profiles multiplied by an integrated spectral weight $A_X$. For coordinate-network outputs $v_j$ at energies $E_j$,
\begin{equation}
 u_j=\operatorname{softplus}(v_j)+\epsilon,
 \qquad
 \widehat Y_j=A_X\frac{u_j}{\max\!\left(\sum_k w_k u_k,\epsilon\right)},
\end{equation}
where $w_k$ are trapezoidal integration weights on the complete physical energy grid and $\epsilon=10^{-8}$. For PhDOS and NWDOS, $A_X$ was fixed at unity, following the approximately unit-area normalization of the source profiles. For 1D INS and eDOS, a linear head acting on $\mathbf z_X$ predicted $\log A_X$, which was clipped to $[-8,12]$ before exponentiation.

Standalone one-dimensional models omitted output layer normalization from the post-pooling projection MLP. In matched supervision, the shared projection retained this normalization, while the eDOS branch used the same full-grid shape-and-area parameterization.

Powder INS, PhDOS, NWDOS and 1D INS used a common PCA transformation fitted to atomic features from their shared training set. For standalone eDOS, PCA was fitted separately using its own training-set atomic features. Matched-supervision models used the PCA transformation fitted to the powder-INS training set, with projected-feature standardization statistics recalculated on the matched training cohort.

Alternative encoder--decoder implementations, the fixed-grid control and the encoder-free single-structure Ne fit are described in Supplementary Note~\ref{note:f1}.

\subsection{Model training and adaptation}\label{sm:training}
The pretrained structural encoder and newly initialized attention-pooling, post-pooling projection, coordinate-network and conditioning modules were jointly optimized for global powder-INS prediction. Reference intensities were standardized using a single mean and standard deviation calculated over all grid values in the training spectra. Training minimized the SmoothL1 loss averaged over sampled structure--coordinate pairs, with a transition parameter of 1 in standardized intensity units. Each minibatch comprised eight structures evaluated at a shared set of 4,096 coordinates sampled uniformly without replacement from the native reference grid. A new coordinate subset was drawn for each minibatch.

Parameters were optimized using AdamW~\cite{loshchilov2019adamw} with a weight decay of $10^{-6}$. Initial learning rates were $10^{-4}$ for the newly initialized modules, $5\times10^{-7}$ for the final two encoder blocks and $5\times10^{-8}$ for the remaining encoder parameters. Gradients were accumulated over two minibatches and clipped to a global $L_2$ norm of 1 before each update. A three-epoch learning-rate warmup was followed by cosine decay to 5\% of the initial rates. Training lasted 60 epochs, with validation over the complete reference grid every five epochs. The checkpoint with the lowest validation mean squared error (MSE) on the reference intensity scale was selected.

Each standalone one-dimensional model was initialized from the pretrained EquiformerV3 encoder, with newly initialized attention-pooling, post-pooling projection and spectral-prediction modules, and trained end to end over the complete energy grid. The spectral objective was the mean squared difference between $\log(1+\widehat Y)$ and $\log(1+Y)$, normalized by the global variance of the log-transformed training targets. For 1D INS and eDOS, a mean squared error in log-integrated spectral weight was added with a coefficient of 0.05; numerical safeguards are specified in Supplementary Note~\ref{note:f2}. Optimization followed the powder-model settings, with all newly initialized modules, including the integrated-weight head where present, assigned the $10^{-4}$ initial learning rate. Training lasted up to 60 epochs, with complete-grid validation after each epoch and early stopping after 12 epochs without improvement. The checkpoint with the lowest variance-normalized validation log-MSE was selected, excluding the integrated-weight auxiliary term.

\phantomsection\label{sm:adaptation}
For strain adaptation, configurations at $\varepsilon=\{-2,-1,0,1,2\}\%$ formed the training set for each compound, whereas those at $\pm1.5\%$ and $\pm0.5\%$ formed the validation and withheld sets, respectively, giving 20 training, eight validation and eight withheld configurations across the four compounds. Composition adaptation used the 15 compositions combining $x=\{0,0.5,1\}$ with $y=\{0,0.25,0.5,0.75,1\}$ for training, while the remaining 30 compositions were withheld.

For the full-spectrum strain-response comparison in Fig.~\ref{fig:3}a, we adapted a model using native-grid reference spectra. A separate model trained on broadened and resampled reference spectra was used for the ridge-shift and gain/loss analyses, together with the strain-response sampling-budget evaluation. Each model was shared across the four compounds, with identical training, validation and withheld partitions. Both strain models and the composition model were independently initialized from the globally trained powder-INS model and fine-tuned for 100 epochs. Adaptation updated all trainable parameters using SmoothL1 loss on absolute spectra, while the global feature transformations and intensity-standardization statistics remained fixed. The optimizer, module-specific learning rates, weight decay, gradient clipping and warmup/cosine-decay schedule followed the global training protocol, with 16,384 sampled coordinates per structure and gradients accumulated over five structures per update.

For each strain model, checkpoints were selected by minimizing mean validation response N-L2 among those whose mean validation absolute-spectrum N-L2 was no greater than 1.10 times its minimum across checkpoints. The composition model used the final-epoch checkpoint and was retained unchanged for size transfer and local spectral queries. Details of the resampled targets and strain-selection tie-breaks are provided in Supplementary Note~\ref{note:f3}.

\phantomsection\label{sm:matched-training}
For the matched INS-only, eDOS-only and joint-supervision experiments, powder-INS and eDOS records were paired by material identifier, using the powder-INS atomic structures and retaining their original training, validation and test assignments. All three supervision conditions used this common cohort of 5,405 training, 719 validation and 649 test materials.

Within each training seed, the three supervision conditions used identical initial parameters for the pretrained encoder and the newly initialized pooling, post-pooling projection and task-specific prediction modules. The Initialization baseline comprised the encoder, pooling and projection modules before spectroscopic training. For joint supervision, a shared material representation conditioned separate INS and eDOS coordinate networks. Models were trained end to end using the corresponding spectral objectives defined above; joint supervision used their sum, including the eDOS integrated-weight term. The INS objective was evaluated at 4,096 sampled coordinates per structure, whereas the eDOS objective used the complete energy grid. Loss-normalization statistics were calculated from the common training cohort.

The matched-supervision models followed the powder-model optimization settings and were trained for 60 epochs, with complete-grid validation every five epochs. Checkpoints were selected by the lowest validation value of the corresponding training objective, including the eDOS integrated-weight term for eDOS-only and joint supervision. Each supervision condition was trained with three random seeds while keeping the data partitions, hyperparameters and checkpoint-selection rules fixed. Implementation details of the matched initialization are provided in Supplementary Note~\ref{note:f4}.

\subsection{Spectral evaluation and analysis}\label{sm:fidelity-metrics}
The global powder-INS model was evaluated over the complete native $Q$--$E$ grid after restoring predictions to the reference intensity scale and clipping negative intensities to zero. PCC was calculated between the vectorized prediction and reference for each material, and N-L2 was defined as
\begin{equation}
 \mathrm{N\!-\!L2}_i=\frac{\|\widehat S_i-S_i\|_2}{\|S_i\|_2},
\end{equation}
where $S_i$ and $\widehat S_i$ contain the reference and predicted grid intensities for material $i$. Population summaries were obtained by aggregating these per-material scores, without spectrum-specific intensity rescaling. Numerical denominator safeguards used $\epsilon=10^{-12}$.

For PhDOS, NWDOS, 1D INS and eDOS, validation N-L2 was calculated over each complete energy grid against the Gaussian-smoothed targets defined above, as $\sqrt{\langle(\widehat Y-Y)^2\rangle/\max[\langle Y^2\rangle,\epsilon]}$, where brackets denote an average over energy points. Absolute-spectrum fidelity for the strain- and composition-adapted models was evaluated using PCC and N-L2 on the corresponding reference grids listed in Supplementary Table~\ref{tab:si-protocol-map}.

Strain-response fidelity was evaluated from the signed differences $\Delta S_\varepsilon=S_\varepsilon-S_0$, with predicted and reference spectra each using their own unstrained baseline. PCC and N-L2 were applied over the complete evaluation grid, retaining both signs and using the reference difference-spectrum norm in the N-L2 denominator.

The integrated intensity, energy centroid and energy-distribution width were obtained from the $Q$-integrated energy profile, $P(E)=\int S(Q,E)\,dQ$, as
\begin{equation}\label{eq:spectral-moments}
\begin{aligned}
 I&=\int P(E)\,dE,\\
 \mu_E&=\frac{1}{I}\int E P(E)\,dE,\\
 \sigma_E&=\left[\frac{1}{I}\int(E-\mu_E)^2P(E)\,dE\right]^{1/2}.
\end{aligned}
\end{equation}
These quantities were evaluated over the complete reference domain using trapezoidal quadrature on the physical $Q$ and $E$ coordinates, without additional smoothing. The same operations were applied to reference and predicted spectra. Intensity error was measured as $|\widehat I-I|/(I+\epsilon)$, whereas centroid and width errors were absolute differences. Here $\sigma_E$ is the standard deviation of the intensity-weighted energy distribution.

Spectral shape was compared using the Wasserstein-1 distance between area-normalized energy profiles~\cite{kong2022mat2spec,vallender1974wasserstein},
\begin{equation}\label{eq:energy-wasserstein}
 W_1=\int\left|\widehat F(E)-F(E)\right|\,dE,
\end{equation}
where $F$ and $\widehat F$ are the cumulative distributions of the separately area-normalized reference and predicted profiles. The discrete probability weights and cumulative-sum implementation are specified in Supplementary Note~\ref{note:f2}. The centroid, width and $W_1$ were expressed in meV.

Peak recovery was assessed using validation-set $Q$-integrated profiles after Gaussian smoothing with $\sigma=0.5$~meV. Maxima were detected with a minimum separation of 2~meV and a prominence threshold of 5\% of each smoothed profile's dynamic range, and reference and predicted peaks were matched one to one within 3~meV. Precision and recall were calculated from pooled peak counts across materials as the fractions of predicted and reference peaks that were matched, respectively. F1 was calculated as the harmonic mean of pooled precision and recall. Peak-energy MAE was calculated only over matched pairs. Numerical implementation and sensitivity analyses are specified in Supplementary Note~\ref{note:f2}.

\phantomsection\label{sm:queries}
Local spectral quantities were evaluated for the 30 withheld alloy compositions using the unchanged composition-adapted model. Within the selected window $\Omega$, spanning $Q=4$--8~\AA$^{-1}$ and $E\simeq10$--60~meV, the local energy centroid was calculated using Eq.~\eqref{eq:spectral-moments}, with both integrations restricted to $\Omega$. The fraction of window-integrated scattering intensity above $E_*\simeq30$~meV was defined as
\begin{equation}\label{eq:local-upper-band}
 w_{>E_*}=\frac{\int_{\Omega\cap\{E\geq E_*\}}S(Q,E)\,dQ\,dE}
 {\int_\Omega S(Q,E)\,dQ\,dE}.
\end{equation}

For each structure, intensities were queried directly from the coordinate network using its cached material representation at $Q=4,5,6,7,8$~\AA$^{-1}$ and eighteen energy coordinates, giving 90 coordinate evaluations. Seventeen energies were selected at approximately evenly spaced native-grid indices between the points nearest 10 and 60~meV, and the threshold coordinate nearest 30~meV was added. Trapezoidal quadrature was applied to these intensities to estimate both local quantities. Sampling errors were calculated relative to estimates obtained from full-grid model predictions over the same window; both sparse and full-grid estimates were also compared with the corresponding reference-spectrum quantities. Absolute errors were averaged across the 30 withheld compositions and expressed in meV for the centroid and percentage points for the intensity fraction. Sampling fractions were calculated relative to the complete 89,700-point reference grid.

For strain-response analysis, selected powder-intensity ridges were tracked in stored predictions from the resampled-domain strain model and the corresponding reference spectra. Tracking started from zero-strain anchors selected using the training and validation reference spectra and proceeded separately toward tensile and compressive strain. At each step, local energy profiles were obtained by integrating intensities over the ten momentum-grid coordinates within $Q=8\pm0.15$~\AA$^{-1}$, using 25 energy samples within $\pm2$~meV of the preceding peak position. Peak energies were refined by three-point quadratic interpolation using the actual sampled energy coordinates. With strain expressed as its percentage value, $\varepsilon_{\%}=100\varepsilon$, the finite-difference coefficient was
\begin{equation}\label{eq:strain-response-coefficient}
 \alpha=\frac{E_{\mathrm{pk}}(+\Delta\varepsilon_{\%})-E_{\mathrm{pk}}(-\Delta\varepsilon_{\%})}
 {2\Delta\varepsilon_{\%}},
\end{equation}
where $E_{\mathrm{pk}}$ is the tracked ridge energy. Coefficients were expressed in meV per percentage point of biaxial strain and compared between predictions and references using the withheld $\pm0.5\%$ states, with coefficient MAE averaged across the four compounds. Ridge selection and interpolation safeguards are specified in Supplementary Note~\ref{note:f4}.

For each material, strain-induced intensity gains and losses were integrated separately within a fixed momentum--energy window $\Omega_m$,
\begin{equation}\label{eq:local-gain-loss}
 W_\pm(\varepsilon)=\int_{\Omega_m}\max[\pm\Delta S_\varepsilon(Q,E),0]\,dQ\,dE,
\end{equation}
where $W_+$ is the integrated intensity gain and $W_-$ is the magnitude of the integrated intensity loss. Each window covered $Q=8\pm0.15$~\AA$^{-1}$ and $|E-E_{0,m}|\leq4$~meV, where $E_{0,m}$ is the material's zero-strain reference ridge anchor. The same window was used for predicted and reference spectra at all strain states. Integrals were evaluated using trapezoidal quadrature on the physical momentum and energy coordinates. Prediction--reference MAEs and PCCs were calculated across the eight withheld $\pm0.5\%$ strain states. Material-specific anchor energies and local sampling grids are specified in Supplementary Note~\ref{note:f4}.

The strain-response sampling-budget analysis used stored model predictions at zero strain and the validation $\pm1.5\%$ states. The number of energy samples within each local tracking window was varied while retaining the momentum samples, ridge-tracking procedure and peak-interpolation rule described above. For each material, the coefficient obtained at each budget was compared with that calculated using all available energy-grid points within the local tracking windows. Mean absolute deviations were calculated across the four compounds. Sampling fractions were defined as the total number of sampled momentum--energy coordinates across the three strain states divided by the number of points on their complete grids. An equal-budget comparison of regular and random energy sampling is described in Supplementary Note~\ref{note:f4}.

\phantomsection\label{sec:si-f2-definitions}
Validation and test materials were assigned to groups E1--E4 using training-set reference-centroid quartiles, with group errors summarized by the median and interquartile range. Compositional lightness was defined as $L_{\min}=m_{\min}^{-1/2}$, where $m_{\min}$ is the lightest constituent atomic mass in atomic mass units, and its association with N-L2 was quantified using Spearman rank correlation. Gaussian kernel-density contours described the training population in $(L_{\min},\mu_E)$. D1--D4 examples were drawn from approximately equal-sized strata of validation N-L2 for each task.

Element-resolved errors were obtained by averaging over evaluation materials containing each element~\cite{chen2021phdos} and compared with the number of training structures containing that element. Each structure was counted once per constituent element. Spearman correlations were evaluated over all represented elements and after requiring at least ten evaluation materials per element. The primary comparison used test materials; validation analyses also varied the error measure. Group boundaries, atomic-fraction-weighted lightness, H-exclusion and common-centroid controls, and gallery-selection rules are specified in Supplementary Note~\ref{note:f2}.

Pointwise 95\% confidence intervals used the 2.5th and 97.5th percentiles of material- or configuration-level bootstrap distributions. Analysis-specific resampling units and replicate counts are given in Supplementary Notes~\ref{note:f2}--\ref{note:f4}. Intervals describe variation across evaluated materials or configurations, conditional on the fitted models. Global powder-model training variability was assessed separately using the mean and sample standard deviation of aggregate test metrics across three independent runs.

\subsection{Analysis of material representations}\label{sm:geometry}
The 256-dimensional material representations of all 45 alloy configurations were extracted from the trained composition-adapted model and visualized by PCA (Fig.~\ref{fig:3}c). PCA was applied to the mean-centered representations, including both training and withheld compositions, and the first two principal components were used for visualization without further rescaling. Lines connect adjacent compositions along the Mo--W and S--Se substitution directions. Distance-based analyses used the original, uncentered 256-dimensional representations.

For the cross-material comparison, neighborhoods were constructed from material representations obtained at initialization and after INS, eDOS or joint supervision under the matched training protocol. Neighbors were ranked by cosine distance,
\begin{equation}
 d_z(i,j)=1-\frac{\mathbf z_i\cdot\mathbf z_j}
 {\|\mathbf z_i\|_2\|\mathbf z_j\|_2}.
\end{equation}
The BaCl$_2$ illustration used the 719-material validation cohort (Fig.~\ref{fig:4}b), whereas population analyses used the 649 held-out test materials (Fig.~\ref{fig:4}c). For each query material $i$, all other materials within its cohort were ranked by increasing $d_z(i,j)$, with the first $k$ defining the neighbor set $N_k^z(i)$. Candidate populations were held fixed across models, and each representation-selected neighbor set was evaluated using both INS and eDOS reference spectra.

Reference spectra were log-transformed, and the transformed intensities were normalized by their root-mean-square value before calculating pairwise spectral distances:
\begin{equation}
 \begin{aligned}
 t_i&=\log[1+\max(S_i,0)],\\
 u_i&=\frac{t_i}{\sqrt{\langle t_i^2\rangle}+10^{-12}},\qquad
 d_{\mathrm{ref}}(i,j)=\sqrt{\langle(u_i-u_j)^2\rangle},
 \end{aligned}
\end{equation}
Here $S_i$ denotes the vectorized reference spectrum of material $i$, and brackets denote an equally weighted average over grid entries. Distances were evaluated over the complete INS grid and the unsmoothed eDOS grid used for the matched-supervision comparison. The same reference-spectral distances were used to evaluate all representation models.

Neighborhood enrichment quantified the fractional reduction in mean reference-spectral distance relative to uniform random-neighbor selection. For $N$ evaluated materials, enrichment was defined as $\mathcal E(k)=1-R(k)$, where
\begin{equation}
 R(k)=\frac{\displaystyle\frac1N\sum_i\frac1k\sum_{j\in N_k^z(i)}d_{\mathrm{ref}}(i,j)}
 {\displaystyle\frac1N\sum_i\frac1{N-1}\sum_{j\ne i}d_{\mathrm{ref}}(i,j)}.
\end{equation}
The numerator averages the reference-spectral distances from each query to its representation-selected neighbors. The denominator averages distances to all eligible non-self candidates, giving the exact expectation for uniform random-neighbor selection. Both averages weight query materials equally. The primary analysis used $k=5$, with $k=10$ and 20 used to examine dependence on neighborhood size.

Agreement between representation-based and reference-spectral neighbor sets was quantified by exact-neighbor recall, $\mathrm{Recall@}k=N^{-1}\sum_i|N_k^z(i)\cap N_k^{\mathrm{ref}}(i)|/k$, where $N_k^{\mathrm{ref}}(i)$ contains the $k$ materials with the smallest reference-spectral distances to query $i$, drawn from the same candidate population as $N_k^z(i)$. Recall was calculated separately for INS and eDOS, with a uniform random-neighbor expectation of $k/(N-1)$.

Neighbor replacement was analyzed for the five-neighbor sets by comparing each trained model with its corresponding initialization. For each query whose neighbor set changed, gained neighbors were those present only after training, and lost neighbors were those present only at initialization. We calculated the mean reference-spectral distance to gained neighbors minus that to lost neighbors and averaged these differences across affected queries. The same gained and lost sets were evaluated against both INS and eDOS references; negative values indicate spectrally closer replacements.

Within each training seed, 95\% confidence intervals were estimated from the 2.5th and 97.5th percentiles of 2,000 bootstrap resamples of query materials. Each resample used the same query indices in the numerator and denominator of $R(k)$, and model contrasts used paired query resampling. Candidate populations and neighbor sets were held fixed. Supervision-induced changes were measured relative to the corresponding initialization for each seed. Figure~\ref{fig:4}b,c reports one matched training run, with two additional runs used to assess training variability, summarized by the mean and sample standard deviation across all three runs. Seed-resolved results and neighborhood-size controls are provided in Supplementary Note~\ref{note:f4}.

\subsection{Computational benchmarking}\label{sm:timing}
We benchmarked complete powder-INS spectrum generation from the same relaxed six-atom WSe$_2$ cell at $-0.5\%$ biaxial strain using DFT-based and EquiformerV3 MLFF-based workflows and direct SpecSCoNe prediction. Both force-based workflows evaluated the same three displaced 96-atom supercells and used identical Phonopy and OCLIMAX settings. The reported DFT time combined the recorded wall times of the three force calculations with the subsequent spectral post-processing time. For the MLFF workflow, force evaluation and force-constant construction were timed together in three measurements following one warmup run; their median was combined with a single measured post-processing run.

Direct SpecSCoNe prediction used the strain-adapted model trained on broadened and resampled spectra, evaluated over its complete $300\times300$ grid. Starting from the structure held in memory and the loaded model, each timed prediction included transfer of atomic inputs and query coordinates to the GPU, graph construction, structural encoding, full-grid spectral evaluation and transfer of the predicted spectrum back to the CPU. The reported time was the median of ten measurements after three warmup runs, with GPU synchronization immediately before and after each timed prediction. Measurements used one NVIDIA RTX PRO 6000 Blackwell GPU and four CPU threads. Reference-workflow hardware, scattering settings and stage-resolved timings are provided in Supplementary Note~\ref{note:f4}.

For the 12--300-atom alloy configurations, reference-generation times were recorded separately for MACE force evaluation and force-constant construction, phonon calculations and input conversion, and OCLIMAX scattering calculations. OCLIMAX settings were held fixed across sizes, including coherent and incoherent scattering with \texttt{TASK=1}, \texttt{MAXO=10} and \texttt{CONV=2}. Phonon meshes were reduced as the supercell dimensions increased to maintain approximately constant reciprocal-space sampling resolution. The unchanged composition-adapted SpecSCoNe model was evaluated over the complete $300\times299$ grid, with five timed predictions after three warmup runs per configuration. Each prediction included atomic-input transfer, graph construction, structural encoding and full-grid evaluation, with GPU synchronization at stage boundaries. Query coordinates were preloaded onto the GPU, and transfer of the predicted spectrum to the CPU occurred after timing. At each size, reference times and configuration-level median prediction times were summarized by their median and range across the two configurations.

Sensitivity to phonon sampling and the scattering approximation was assessed using the same MACE force constants for a relaxed 300-atom alloy configuration at $(x,y)=(0.5,0.5)$. One control reduced the phonon mesh from $7\times7\times1$ to $2\times2\times1$ while retaining \texttt{TASK=1}; the other retained the original mesh and used the incoherent approximation (\texttt{TASK=0}). Temperature, broadening, intensity normalization and output coordinates were unchanged. Timings covered OCLIMAX execution only. Spectral differences were quantified relative to the original \texttt{TASK=1} calculation using N-L2 for the complete powder maps, $Q$-integrated energy profiles and constant-$Q$ profiles, without additional intensity rescaling. Execution settings and detailed comparisons are provided in Supplementary Note~\ref{note:f4}.

For the 11,532-atom configuration, the same composition-adapted SpecSCoNe model evaluated all 89,700 spectral coordinates using 32-bit floating-point precision (FP32) on a single NVIDIA RTX PRO 6000 Blackwell GPU with four CPU threads. The atomic graph and material representation were recomputed for every prediction. Each of three independently launched processes performed three warmup runs followed by five timed predictions, and the reported runtime was the median of the three within-process medians. Timing followed the alloy-size benchmark, ending before transfer of the predicted spectrum to the CPU. Memory-allocation settings and separate memory-profiling measurements are described in Supplementary Note~\ref{note:f3}.

%% file: sections/declarations.tex


\section*{Data availability}

The data supporting the findings of this study are derived from publicly available computational datasets and simulations performed as described in the Methods. Additional data are available from the corresponding authors upon reasonable request.

\section*{Code availability}

The source code required to reproduce the results presented in this work will be made publicly available through a GitHub repository with a permanent Zenodo archive upon acceptance of the manuscript.


\section*{Competing interests}

The authors declare no competing interests.

\begin{acknowledgments}
This work is primarily supported by the U.S. Department of Energy, Office of Science, Basic Energy Sciences (BES) under the Genesis Mission BES AI Pathfinder Program, MAIQMag: Multimodal AI for 2D Quantum Magnets.
This research used computational resources of the National Energy Research Scientific Computing Center (NERSC), a US Department of Energy Office of Science User Facility located at Lawrence Berkeley National Laboratory operated under contract DE-AC02-05CH11231, using NERSC award ERCAP0038195.
The authors acknowledge the Texas Advanced Computing Center (TACC) at The University of Texas at Austin for providing computational resources under project code DMR25014 that have contributed to the research results reported within this paper. 
During the preparation of this work, the authors used the large language model ChatGPT by OpenAI in order to refine the language and enhance the readability of this paper. After using this tool, the authors reviewed and edited the content as needed and take full responsibility for all content in this publication.
\end{acknowledgments}

%% file: si_revision.tex
\begingroup
\section*{Supplementary Information}
\setcounter{subsection}{0}
\renewcommand{\thesubsection}{\arabic{subsection}}
\renewcommand{\thesubsubsection}{\thesubsection.\arabic{subsubsection}}
\makeatletter
\def\p@subsection{}
\def\p@subsubsection{}
\makeatother
\renewcommand{\theHsubsection}{supp.\arabic{subsection}}
\renewcommand{\theHsubsubsection}{supp.\arabic{subsection}.\arabic{subsubsection}}
\setcounter{figure}{0}
\setcounter{table}{0}
\setcounter{equation}{0}
\renewcommand{\thefigure}{S\arabic{figure}}
\renewcommand{\thetable}{S\arabic{table}}
\renewcommand{\theequation}{S\arabic{equation}}
\renewcommand{\theHfigure}{supp.\arabic{figure}}
\renewcommand{\theHtable}{supp.\arabic{table}}
\renewcommand{\theHequation}{supp.\arabic{equation}}

\noindent Table~\ref{tab:si-protocol-map} summarizes the model instances, reference populations and observation domains. Supplementary Notes~\ref{note:f1}--\ref{note:f4} provide additional results and experiment-specific settings.
\begin{table*}[!htbp]
\caption{\textbf{Models, reference populations and observation domains.} Counts refer to materials or configurations. The native powder grid spans approximately $Q=0.05$--14.95~\AA$^{-1}$ and $E=0.50$--149.995~meV.}
\label{tab:si-protocol-map}
\centering\small
\begin{tabularx}{\linewidth}{>{\raggedright\arraybackslash}p{0.17\linewidth}YYY}
\toprule
Study & Model and training population & Evaluation and reference & Observation domain\\
\midrule
Global powder (Figs.~\ref{fig:1},\ref{fig:2}) & Fully fine-tuned encoder and decoder; 8,021 train, 1,001 validation & 1,001 test; reference powder INS & Native $300\times299$; raw-intensity fidelity\\
Implementation / raster controls & Same global-powder split; alternative encoders/decoders or rank-44 raster & 1,001 validation materials; selected by complete-grid raw MSE & Native powder grid; implementation and fidelity controls\\
Single-crystal Ne fitting & Crystal-specific latent code and coordinate network fitted jointly & Reconstruction of the Ne reference used for fitting & $26^3\times750$; four-dimensional representation capacity\\
Standalone 1D (Fig.~\ref{fig:2}) & Four separate models; 8,021 vibrational or 5,421 electronic train & 1,001 vibrational or 676 electronic validation; smoothed reference targets & 2,999 points, $\sim$0.1--300~meV; eDOS 2,001 points, $[-10,10]$~eV\\
Shared strain (Fig.~\ref{fig:3}a) & One four-endmember adapted model; 20 train, 8 validation & Eight withheld $\pm0.5\%$ states; DFT force constants/OCLIMAX & Native $300\times299$; structural response\\
Composition / size (Fig.~\ref{fig:3}b--d) & Composition-adapted model; 15 compositions, 12 atoms each & 30 withheld compositions; two compositions at five cell sizes; specialized MACE/OCLIMAX & Native $300\times299$; composition and size transfer\\
Single-GPU capacity (Fig.~\ref{fig:3}e) & Same composition-adapted model; no further adaptation & Independently randomized configuration at $(x,y)=(0.25,0.625)$; 11,532 atoms & Native $300\times299$; inference time and memory use\\
Matched supervision (Fig.~\ref{fig:4}a--c) & Initialization, INS, eDOS and joint supervision; three seeds; 5,405 train, 719 validation & 649 common test materials; reference INS/eDOS distances & Native powder and $[-10,10]$~eV eDOS; full 256D latent neighborhoods\\
Composition queries (Fig.~\ref{fig:4}d,f) & Same composition-adapted model as Fig.~\ref{fig:3}b--d & Same 30 withheld compositions; coordinate queries using cached material representations & Local $Q=4$--8~\AA$^{-1}$, $E\simeq10$--60~meV\\
Strain readout (Fig.~\ref{fig:4}e,f) & Separate shared-family adapted model; 20 train, 8 validation; distinct from Fig.~\ref{fig:3}a & Eight withheld states for coefficients; validation states for budget; subsampled stored spectra & $300\times300$, $Q=0.05$--9.276~\AA$^{-1}$, $E=0.5$--44.575~meV\\
Workflow timing (Fig.~\ref{fig:4}g) & Strain-readout model for WSe$_2$; composition-adapted model for alloy-size series & One relaxed WSe$_2$ case and two alloy configurations per size & Strain-readout grid for WSe$_2$; native grid for alloy series; spectrum-generation time\\
\bottomrule
\end{tabularx}
\end{table*}

\clearpage

\subsection{Spectral fidelity across implementations and observation domains}\label{note:f1}
\subsubsection{Coordinate-based prediction versus a fixed-grid decoder}\label{sm:raster}
To examine whether coordinate-based evaluation affects reconstruction on the native reference grid, we compared the modulated SIREN with a low-rank fixed-grid decoder of comparable parameter count.

Across 1,001 validation materials, the coordinate model achieved a mean N-L2 of 0.230, compared with 0.264 for the fixed-grid model, while reducing the error for 69.0\% of materials (Supplementary Fig.~\ref{fig:si-field-raster}). The paired mean reduction was 0.03467, with a 95\% confidence interval of 0.02965--0.03999 from 5,000 paired material-bootstrap resamples.

\begin{figure}[!htbp]
\centering
\includegraphics[width=0.97\textwidth]{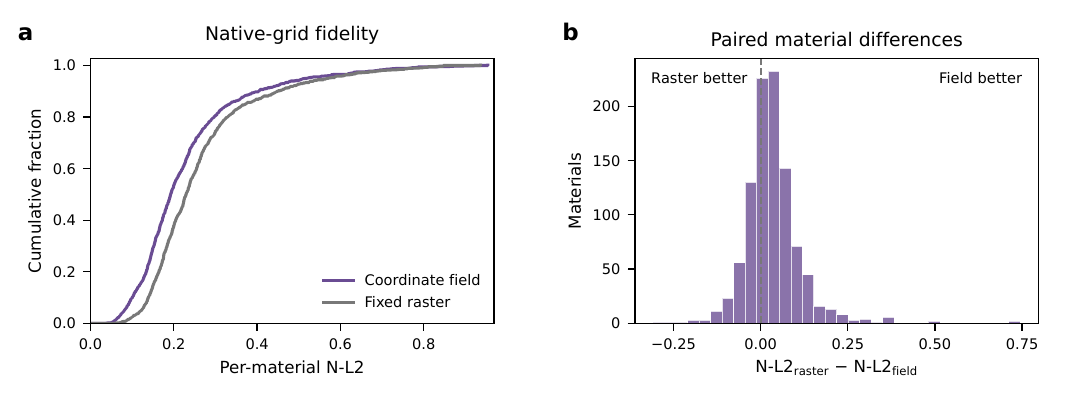}
\caption{\textbf{Native-grid fidelity of coordinate-based and fixed-grid prediction.} \textbf{a,} Per-material N-L2 distributions for 1,001 validation materials on the $300\times299$ grid. \textbf{b,} Paired differences, N-L2$_{\mathrm{raster}}-$N-L2$_{\mathrm{field}}$; positive values favor the field.}
\label{fig:si-field-raster}
\end{figure}

\paragraph{Decoder implementation and matched training.}
The fixed-grid decoder represented standardized intensity at native pixel $p$ as
\[
g_p(\mathbf z_X)=b_p+\sum_{r=1}^{44}c_r(\mathbf z_X)B_{rp},
\]
where $B_{rp}$ are learned spectral basis values, $b_p$ is a learned pixelwise bias, and a $256\rightarrow256\rightarrow44$ MLP with a SiLU hidden activation predicts the material-dependent coefficients $c_r$. A rank of 44 gave 4,113,600 decoder parameters, compared with 4,336,129 for the coordinate network. The basis and bias were defined on the native $300\times299$ grid and learned jointly with the coefficient network.

Both models shared the pretrained EquiformerV3 initialization, feature transformations, pooling architecture, material partitions and intensity normalization. End-to-end training followed the global powder protocol in \hyperref[sm:training]{Methods}, including sampled-coordinate supervision and complete-grid validation. Both selected checkpoints corresponded to epoch 60.

\subsubsection{Reconstruction across encoder--decoder implementations}\label{sm:alternatives}
To assess the dependence of reconstruction fidelity on the network implementation, we evaluated four structural encoders: MACE~\cite{batatia2022mace}, SevenNet~\cite{park2024sevennet,kim2026sevennetomni}, EquiformerV3~\cite{liao2026equiformerv3} and EquFlash~\cite{samsungdsEquFlash}. Each encoder was paired with three spectral decoders: FiLM-conditioned SIREN~\cite{sitzmann2020siren,perez2018film}, multiplicative decoder adapted from a learnable-kernel formulation~\cite{ni2025dual}, and wavelet-based decoder adapted from WIEN-INR~\cite{ni2026wien}. All configurations used the same material partitions and reference spectra.

High reconstruction fidelity was retained across the twelve implementations: on the 1,001 validation materials, mean per-material PCC ranged from 0.937 to 0.946 and MSE from 0.5706 to 0.7135 on the common reference-intensity scale (Supplementary Table~\ref{tab:si-implementations}). Decoder capacities and optimization settings are specified below.

\begin{table}[!htbp]
\caption{\textbf{Alternative encoder and spectral-decoder implementations.} Complete-grid results for 1,001 validation materials. MSE uses the common reference-intensity scale; PCC is the mean per-material correlation.}
\label{tab:si-implementations}
\centering\small
\begin{tabular}{lrrrrrr}
\toprule
 & \multicolumn{2}{c}{SIREN} & \multicolumn{2}{c}{Multiplicative} & \multicolumn{2}{c}{WIEN-INR}\\
\cmidrule(lr){2-3}\cmidrule(lr){4-5}\cmidrule(lr){6-7}
Encoder & MSE & PCC & MSE & PCC & MSE & PCC\\
\midrule
MACE & 0.6869 & 0.9406 & 0.7046 & 0.9366 & 0.7135 & 0.9395\\
SevenNet & 0.6335 & 0.9429 & 0.6588 & 0.9398 & 0.6421 & 0.9429\\
EquiformerV3 & 0.5706 & 0.9460 & 0.6008 & 0.9437 & 0.5952 & 0.9458\\
EquFlash & 0.6201 & 0.9446 & 0.6445 & 0.9417 & 0.6326 & 0.9450\\
\bottomrule
\end{tabular}
\end{table}

\paragraph{Implementation settings.}
Fixed training-derived projections and standardization transformed each encoder's atomic features, followed by attention pooling and projection to a 256-dimensional material representation. Each configuration was trained once for 60 epochs, jointly updating the encoder, pooling, projection and decoder. Checkpoints minimized complete-grid validation MSE evaluated every five epochs.

The SIREN implementation is defined in \hyperref[sm:encoder]{Methods}~\cite{sitzmann2020siren,perez2018film}. The multiplicative decoder adapted a learnable-kernel formulation, combining a base field with bounded, coordinate-dependent modulation~\cite{ni2025dual}:
\[
\widehat y(\mathbf z_X,\boldsymbol{\xi})
=b(\mathbf z_X^{(b)},\boldsymbol{\xi})
\left[1+\tanh k(\mathbf z_X^{(k)},\boldsymbol{\xi})\right],
\]
where $\widehat y$ denotes standardized intensity and the two jointly trained FiLM--SIREN branches were conditioned on separate 128-dimensional halves of the material representation. The wavelet-based decoder adapted WIEN-INR to predict approximation and detail coefficients of the standardized powder-INS map, followed by inverse wavelet reconstruction~\cite{ni2026wien}. Our implementation used a single-level bior2.2 decomposition with periodization. An auxiliary network provided a pointwise modulation of a learned approximation-to-detail projection, which was added to the direct detail prediction in place of the original local-neighborhood enhancement.

Unless specified below, the optimizer, minibatch size, gradient accumulation and learning-rate schedule followed the main powder-model protocol (Methods). Decoder-specific settings were as follows.

\begin{center}\small
\begin{tabularx}{\linewidth}{>{\raggedright\arraybackslash}p{0.29\linewidth}YYY}\toprule
Decoder setting & SIREN & Multiplicative & Adapted WIEN-INR\\\midrule
Sinusoidal-layer width & 512 & 384 per branch & 384 for coefficient branches; 128 for auxiliary modulation\\
Hidden layers after the input layer & 4 & 4 per branch & 3 per branch\\
Non-encoder learning rate & $10^{-4}$ & $3\times10^{-5}$ & $10^{-4}$\\
Gradient clipping norm & 1 & 0.5 & 1\\
Training target & Standardized intensity & Standardized intensity & Wavelet coefficients of standardized intensity\\
Sampled locations per minibatch & 4,096 & 4,096 & 4,096 per coefficient grid\\\bottomrule
\end{tabularx}
\end{center}
For the wavelet-based decoder, SmoothL1 losses were averaged separately over the approximation coefficients and the three-channel detail coefficients, and the two resulting losses were given equal weight. The auxiliary modulation was the central component of a nine-output softmax generated by the auxiliary FiLM--SIREN.

For each FiLM--SIREN branch, first-layer weights were initialized uniformly in $[-1/d_{\mathrm{in}},1/d_{\mathrm{in}}]$, whereas subsequent sine layers and the final linear output layer used $[-\sqrt{6/d_{\mathrm{in}}}/\omega_0,\sqrt{6/d_{\mathrm{in}}}/\omega_0]$, where $d_{\mathrm{in}}$ is the input dimension of the respective layer and $\omega_0=30$~\cite{sitzmann2020siren}. Biases in these layers were initialized to zero. The conditioning-MLP linear layers and the WIEN approximation-to-detail projection retained the default PyTorch initialization.

Upper/lower encoder learning rates were $2\times10^{-6}/3\times10^{-7}$ (MACE), $10^{-6}/10^{-7}$ (SevenNet) and $5\times10^{-7}/5\times10^{-8}$ (EquiformerV3 and EquFlash). Predictions were transformed back to the reference intensity scale and clipped at zero before evaluation. MSE was averaged over all materials and grid points, whereas PCC was calculated separately for each material and then averaged. The direct coordinate-query analyses used the SIREN implementation.

\subsubsection{Four-dimensional single-crystal spectrum fitting}\label{sm:4d-capacity}
To assess four-dimensional representation capacity independently of structural encoding, we jointly fitted a crystal-specific latent code and a Fourier--sine coordinate network to the Ne reference in $(Q_x,Q_y,Q_z,E)$. The reference contained $26^3$ momentum points and 750 energies spanning approximately 0.2--150~meV. Reconstruction was evaluated on the complete reference grid used for fitting.

In reciprocal-space sections summed over 0--10~meV, the fitted field reproduces the principal ring- and arc-like intensity patterns, while the momentum-averaged profile recovers the dominant low-energy features (Supplementary Fig.~\ref{fig:si-capacity}). PCC evaluated over individual grid points was 0.822 across the complete four-dimensional domain and 0.848 within 0--10~meV.

The full-energy profile reveals residual fitted intensity above 10~meV, where the reference is zero. Because positive intensity occupies only 2.47\% of the reference grid, this broad background exposes fitting errors that are less apparent in the low-energy sections. The two energy ranges therefore distinguish recovery of localized scattering features from reproduction of the surrounding background.

\begin{figure}[!htbp]
\centering
\includegraphics[width=0.97\textwidth,height=0.62\textheight,keepaspectratio]{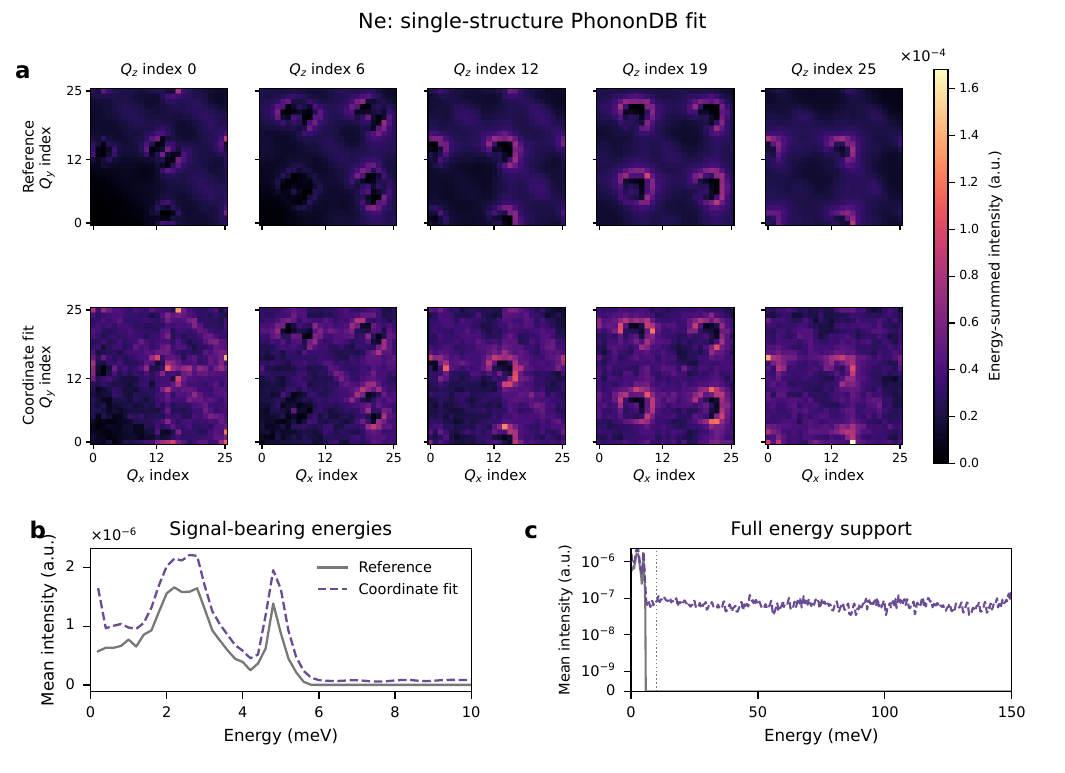}
\caption{\textbf{Single-structure fitting of a sparse single-crystal Ne spectrum.} \textbf{a,} Ne reference and coordinate-network fit at five native $Q_z$ sections, summed over 0--10~meV with a shared linear intensity scale. Momentum axes are grid indices. \textbf{b,c,} Momentum-averaged intensity over the signal-bearing and full energy ranges; c uses a symmetric-log scale.}
\label{fig:si-capacity}
\end{figure}
\paragraph{Fitting and evaluation details.}
The Ne reference (\texttt{mp-111}) was drawn from the separate PhononDB-derived single-crystal INS collection described in \hyperref[sm:data]{Methods}. Coordinates were independently rescaled to $[-1,1]$. The trainable latent code conditioned a modulated sine branch~\cite{sitzmann2020siren} and a Fourier-feature residual branch~\cite{tancik2020fourier}. Log-transformed intensities, $y=\log(1+S)$, were standardized using their full-reference mean and sample standard deviation, with a standard-deviation floor of $10^{-12}$. AdamW minimized the squared error in these standardized targets. Half the sampled points were drawn uniformly from the complete domain, while the other half were drawn from the 0--10~meV window.

Predictions were transformed back to the reference intensity scale by reversing standardization and applying $\operatorname{expm1}$. PCC used the unclipped intensities, whereas displayed predictions were clipped at zero. Reciprocal-space maps summed the 50 native energy bins within 0--10~meV without a $\Delta E$ factor and displayed $Q_z$ indices 0, 6, 12, 19 and 25. Reference and fitted maps shared a linear scale from zero to their joint maximum, without spatial smoothing. Energy profiles averaged all $26^3$ momentum points; the full-energy profile used a symmetric-log scale, linear within $10^{-9}$ of zero.

\par\bigskip

\subsection{Spectral fidelity and material-dependent difficulty}\label{note:f2}

\subsubsection{Population fidelity and peak recovery}
Validation and test materials had similar full-grid N-L2 and PCC distributions, with mean N-L2 values of 0.230 and 0.233, respectively (Supplementary Fig.~\ref{fig:si-fidelity-features}a,b). Across three training runs, mean test N-L2 varied by less than 0.0008 (Supplementary Table~\ref{tab:si-f2-seeds}).

\begin{table*}[htbp]
\caption{\textbf{Stability across independent training runs.} Per-material metrics averaged over the same 1,001 test materials. The final row reports the mean and sample standard deviation (SD) across runs.}
\label{tab:si-f2-seeds}
\centering\small
\begin{tabular}{lrr}\toprule
Training run & Mean test N-L2 & Mean per-material PCC\\\midrule
1 & 0.233211 & 0.943304\\
2 & 0.233624 & 0.942381\\
3 & 0.234003 & 0.942558\\
Across runs & $0.233613\pm0.000396$ & $0.942748\pm0.000490$\\
\bottomrule\end{tabular}
\end{table*}

Peak analysis distinguishes the localization of recovered features from the completeness of peak recovery. At 5\% prominence and a 3~meV matching tolerance, validation-set $Q$-integrated profiles gave pooled precision of 0.730 and recall of 0.433, while successfully matched peaks had an energy MAE of 1.12~meV (Supplementary Fig.~\ref{fig:si-fidelity-features}c). Varying prominence from 2.5\% to 10\% at the same tolerance gave recall values of 0.429--0.460, so the lower recovery of individual maxima was not specific to the primary detection threshold (Supplementary Fig.~\ref{fig:si-fidelity-features}d).

\begin{figure}[!htbp]
\centering
\includegraphics[width=\textwidth,height=0.67\textheight,keepaspectratio]{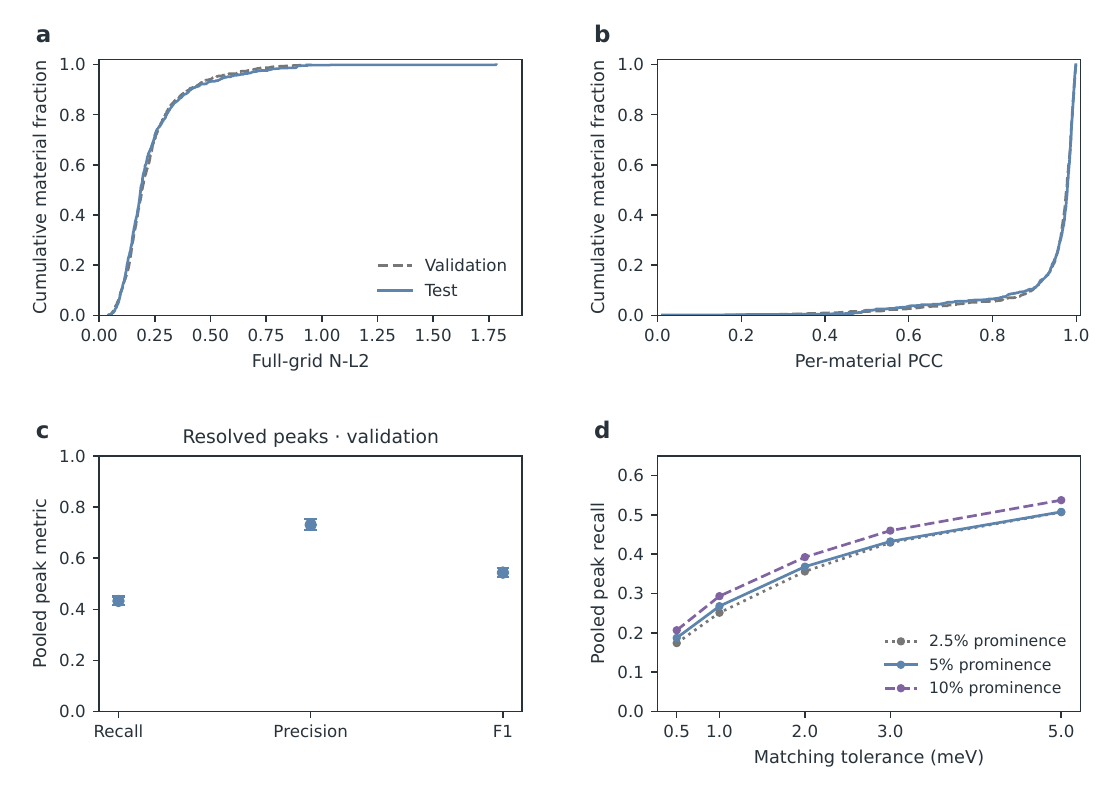}
\caption{\textbf{Population fidelity and peak recovery.} \textbf{a,b,} Full-grid N-L2 and PCC distributions for 1,001 validation and 1,001 test materials (dashed/solid). \textbf{c,} Pooled peak recall, precision and F1 across the 1,001 validation materials at 5\% prominence and 3~meV matching tolerance; bars are 95\% material-bootstrap intervals. \textbf{d,} Recall across prominence and matching-tolerance settings. Peaks are local maxima of Gaussian-smoothed $Q$-integrated profiles.}
\label{fig:si-fidelity-features}
\end{figure}

\subsubsection{Expanded spectral comparisons}
Supplementary Figs.~\ref{fig:si-powder-atlas} and \ref{fig:si-1d-atlas} extend the comparisons to the complete reference domains, including powder energies outside the display range of Fig.~\ref{fig:2}c. Full-validation one-dimensional errors are reported in Supplementary Table~\ref{tab:si-1d-validation}.

\begin{table*}[!htbp]
\caption{\textbf{Validation fidelity across one-dimensional observables.} Full-validation N-L2 against Gaussian-smoothed reference spectra. Gaussian smoothing is specified by its full width at half maximum (FWHM); brackets give the interquartile range (IQR; 25th–75th percentiles).}
\label{tab:si-1d-validation}
\centering\small
\begin{tabular}{lrrrr}
\toprule
Task & Train / validation & FWHM & Mean N-L2 & Median N-L2 [IQR]\\
\midrule
PhDOS & 8,021 / 1,001 & 8 meV & 0.241 & 0.186 [0.134, 0.270]\\
NWDOS & 8,021 / 1,001 & 8 meV & 0.217 & 0.172 [0.114, 0.257]\\
1D INS & 8,021 / 1,001 & 8 meV & 0.195 & 0.163 [0.118, 0.235]\\
eDOS & 5,421 / 676 & 0.8 eV & 0.280 & 0.241 [0.174, 0.356]\\
\bottomrule
\end{tabular}
\end{table*}

\begin{figure}[p]
\centering
\includegraphics[width=\textwidth,height=0.84\textheight,keepaspectratio]{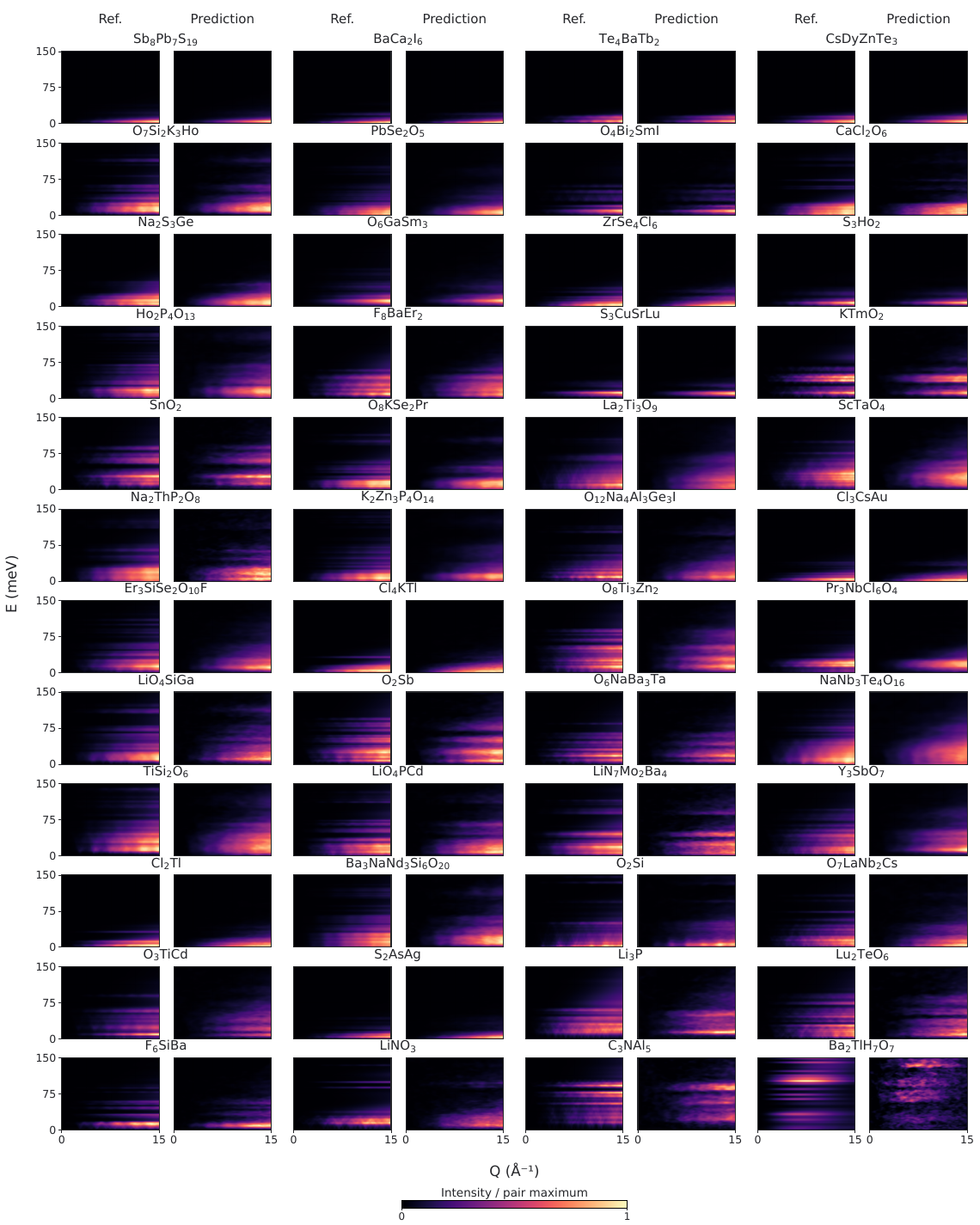}
\caption{\textbf{Powder-INS prediction examples.} Reference and predicted spectra for 48 validation materials over the complete native $Q$--$E$ grid. Examples were randomly selected within successive validation N-L2 quartiles, with 16, 16, 8 and 8 materials, respectively, and are arranged by increasing error. Reference and prediction share a colour scale within each material.}
\label{fig:si-powder-atlas}
\end{figure}

\begin{figure}[p]
\centering
\includegraphics[width=\textwidth,height=0.84\textheight,keepaspectratio]{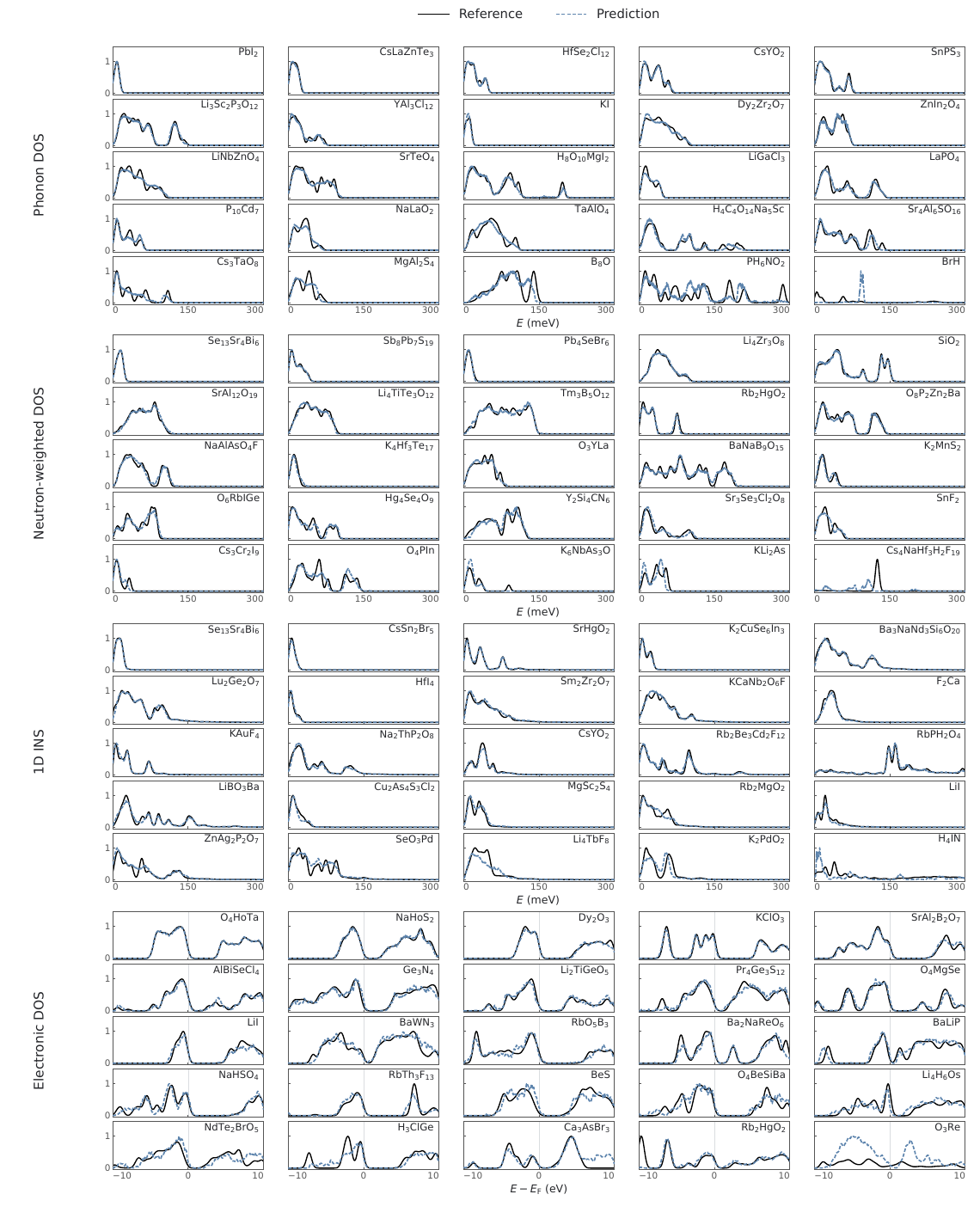}
\caption{\textbf{One-dimensional spectral prediction across four observables.} Reference and predicted PhDOS, NWDOS, 1D INS and eDOS profiles for 25 validation materials per task. Examples span each task's validation-error range and are arranged by increasing full-grid N-L2, from left to right and top to bottom. The complete reference energy grids are shown, using the Gaussian-smoothed targets specified in Methods without additional display smoothing. Reference and prediction share an intensity scale within each comparison.}
\label{fig:si-1d-atlas}
\end{figure}

\clearpage

\subsubsection{Composition-dependent prediction difficulty}
To examine whether the lightness--error association depends on using the lightest constituent, we replaced $L_{\min}$ with atomic-fraction-weighted lightness, $L_{\mathrm{comp}}=\sum_Zx_Zm_Z^{-1/2}$, where $x_Z$ and $m_Z$ are the atomic fraction and mass of element $Z$. Spearman correlations with N-L2 were 0.430 and 0.406 on validation and test, respectively, and remained positive after excluding H-containing materials (0.328 and 0.288). The association thus persists under both a composition-averaged descriptor and removal of the H-containing subset (Supplementary Fig.~\ref{fig:si-material-controls}c).

Elemental occurrence provides a less consistent account of reconstruction difficulty. In validation materials, the correlation between training occurrence and mean N-L2 changed from $-0.207$ across all represented elements to $+0.180$ when each element was required to occur in at least ten validation materials. Replacing N-L2 with raw MSE likewise gave no consistent association between greater occurrence and lower error (Supplementary Table~\ref{tab:si-f2-composition-controls}), extending the test-set comparison in Fig.~\ref{fig:si-material-controls}b across error measures and group-size thresholds.

\begin{figure}[!htbp]
\centering
\includegraphics[width=\textwidth,height=0.67\textheight,keepaspectratio]{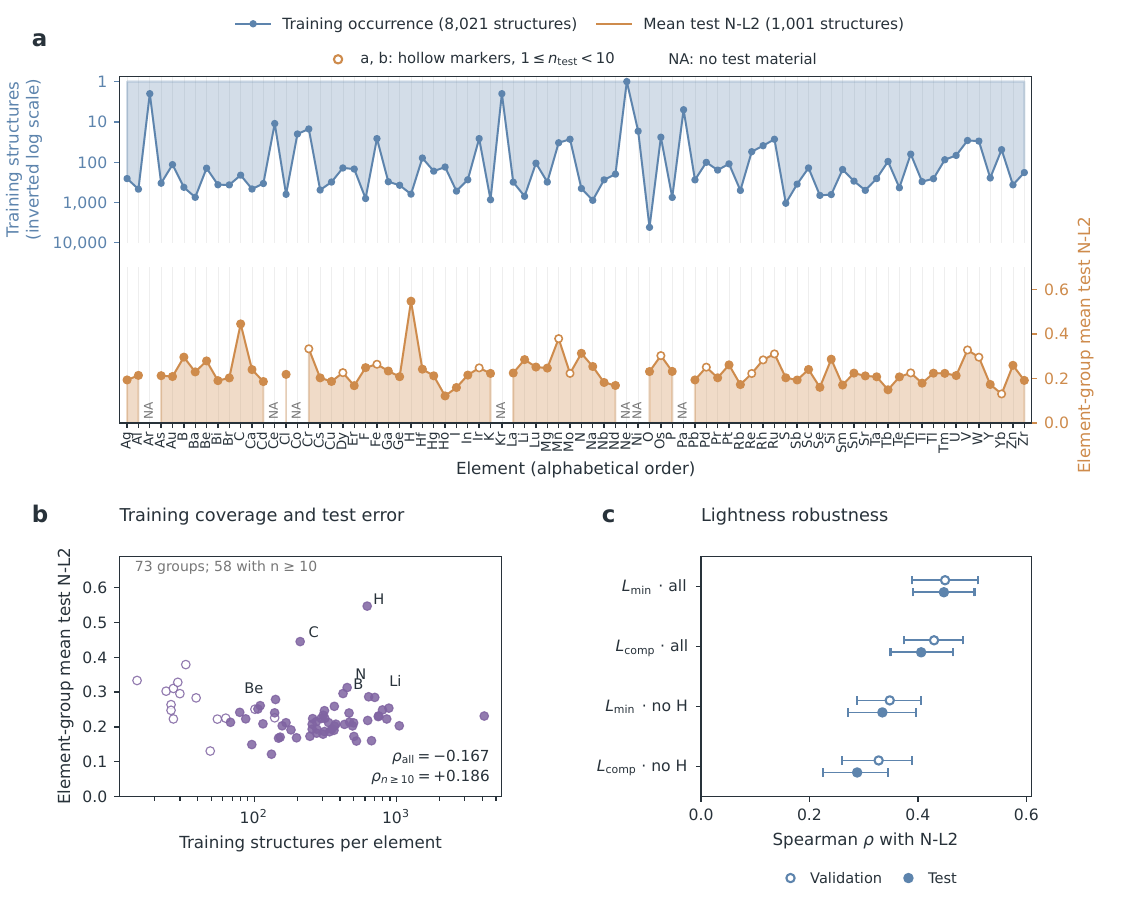}
\caption{\textbf{Element coverage and composition-dependent prediction difficulty.} \textbf{a,} Training occurrence and mean test N-L2 share an alphabetical axis across 80 training elements; upper counts use an inverted logarithmic scale, and lower errors a linear scale. \textbf{b,} Training occurrence versus mean test N-L2 for 73 represented elements, with correlations for all groups and those containing at least ten test materials. Open markers in a,b indicate fewer than ten test materials; missing groups are NA. Element groups overlap. \textbf{c,} Spearman associations between N-L2 and minimum-mass or composition-weighted lightness, for all materials and after excluding H-containing materials. Open/filled markers denote validation/test; bars are pointwise 95\% material-bootstrap intervals.}
\label{fig:si-material-controls}
\end{figure}

\begin{table*}[!htbp]
\caption{\textbf{Sensitivity of the element-coverage association.} Spearman correlation between training occurrence and mean validation error for each element group, with pointwise 95\% material-bootstrap intervals. Negative values associate greater occurrence with lower error.}
\label{tab:si-f2-composition-controls}
\centering\small
\begin{tabular}{llrrl}\toprule
Error measure & Minimum validation count & Elements & $\rho$ & 95\% interval\\\midrule
Raw MSE & 1 & 77 & 0.004 & $[-0.058,\ 0.201]$\\
Raw MSE & 10 & 59 & 0.328 & $[0.252,\ 0.463]$\\
N-L2 & 1 & 77 & $-0.207$ & $[-0.254,\ 0.017]$\\
N-L2 & 10 & 59 & 0.180 & $[0.062,\ 0.287]$\\
\bottomrule\end{tabular}
\end{table*}

The H-containing and H-free populations also occupy different spectral-energy regimes (Supplementary Fig.~\ref{fig:si-hconditional}). Restricting test materials to reference centroids of 50--70~meV retained a positive H-minus-non-H median N-L2 difference of 0.359 (95\% interval $[0.251,0.559]$; 13 H-containing and 44 H-free materials), but substantial centroid imbalance remained, with a standardized mean difference (SMD) of 2.13. In the 60--80~meV window, this imbalance was smaller (0.40), while the error-difference interval spanned zero ($[-0.434,0.301]$; 50 H-containing and 16 H-free materials). Shared energy support does not imply matched spectral-energy distributions, as the residual centroid imbalance illustrates.

\begin{figure}[!htbp]
\centering
\includegraphics[width=\textwidth,height=0.32\textheight,keepaspectratio]{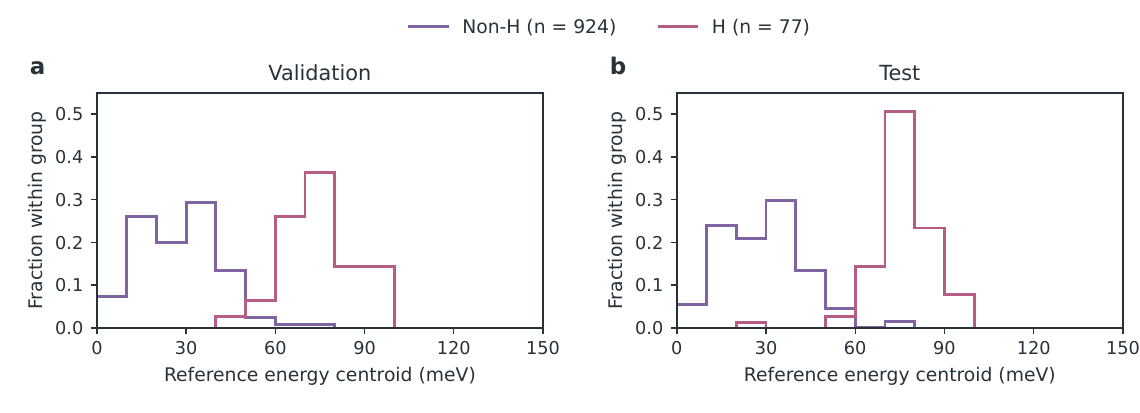}
\caption{\textbf{Different spectral regimes of H-containing and non-H materials.} Separately normalized reference-energy-centroid distributions on \textbf{a,} validation and \textbf{b,} test, using common 10~meV bins. Both splits contain 77 H-containing and 924 non-H materials.}
\label{fig:si-hconditional}
\end{figure}

\subsubsection{Analysis details}\label{sec:si-extra-analysis}
\paragraph{Peak recovery.}
For the peak-recovery analysis defined in \hyperref[sm:fidelity-metrics]{Methods}, Gaussian smoothing used nearest-boundary handling and truncation at $4\sigma$. The minimum peak separation was implemented as $\lceil2.0\,\mathrm{meV}/\operatorname{median}(\Delta E)\rceil$ energy bins. Only interior maxima were detected, and profiles with a smoothed dynamic range no greater than $10^{-12}$ yielded no peaks. One-to-one matching first maximized the number of matches within the prescribed tolerance and then minimized their total energy mismatch, without subgrid fitting. Sensitivity analysis combined prominence thresholds of 2.5, 5 and 10\% with matching tolerances of 0.5, 1, 2, 3 and 5~meV.

The primary settings yielded 1,807 matches from 4,178 reference and 2,474 predicted maxima. Pooled precision and recall had 95\% material-bootstrap intervals of $[0.708,0.754]$ and $[0.416,0.450]$, respectively. Pooled F1 was 0.543 $[0.526,0.562]$, and matched-energy MAE was 1.115~meV $[1.070,1.159]$.

\paragraph{Population controls.}
The training reference-centroid quartiles used for E1--E4 were rounded to 0.1~meV, giving boundaries of 16.7, 30.3 and 39.2~meV. A value on a boundary was assigned to the higher-centroid group.

Training-density contours in $(L_{\min},\mu_E)$ used a Gaussian kernel with Scott's bandwidth and enclosed 50, 80 and 95\% of the estimated probability mass.

The common-centroid windows were selected using validation materials. No 10~meV bin contained at least ten H-containing and ten H-free materials; applying this criterion to 20~meV bins with origins at 0 and 10~meV identified the 50--70 and 60--80~meV windows, which were then retained for test evaluation. Remaining centroid imbalance was quantified by
\[
\mathrm{SMD}=\frac{\overline\mu_H-\overline\mu_{\rm nonH}}
{\sqrt{(s_H^2+s_{\rm nonH}^2)/2}},
\]
where $\overline\mu$ and $s^2$ are the within-window mean and sample variance of the reference energy centroid for each group.

\paragraph{Uncertainty.}
Material-bootstrap resampling schemes and replicate counts were:
\begin{center}\small
\begin{tabularx}{\linewidth}{lYY}\toprule
Analysis & Material resampling & Replicates\\\midrule
Lightness & Separate draws by split & 700\\
Element occurrence & Shared validation draws across overlapping groups & 2,000\\
Hydrogen & Separate H/non-H draws, shared across full/restricted windows & 2,000\\
Peak recovery & Whole materials, retaining all associated peak statistics & 2,000\\\bottomrule
\end{tabularx}
\end{center}
Element training counts remained fixed, and empty resampled groups were omitted; the statistical scope of these intervals is defined in \hyperref[sm:fidelity-metrics]{Methods}.

\paragraph{Gallery selection and scaling.}
For the 48-material powder gallery, 16, 16, 8 and 8 examples were sampled without replacement from the four validation-error quartiles (D1--D4), respectively, using random seed 20260927. For each one-dimensional task, 16 examples---four per quartile at within-quartile rank percentiles of 20, 40, 60 and 80---were supplemented by nine examples at the nearest unused ranks to global percentiles $0,12.5,\ldots,100$. Display scales were shared within each reference--prediction pair; errors were calculated on the reference-data intensity scale.

\paragraph{Numerical safeguards for one-dimensional training.}
For one-dimensional training, the variance normalizing the log-spectral loss was floored at $10^{-12}$, and reference integrals in the 1D INS and eDOS log-integrated-weight losses were floored at $10^{-8}$. The same safeguards applied to matched eDOS supervision.

\paragraph{Discrete Wasserstein-1 calculation.}
For each $Q$-integrated profile, trapezoidal energy weights $w_j^E$ defined discrete masses $p_j=w_j^EP(E_j)/(I+\epsilon)$, with $\epsilon=10^{-12}$ and $I=\sum_jw_j^EP(E_j)$. The cumulative masses were $F_j=\sum_{r\leq j}p_r$, with the same construction for the prediction. We evaluated $W_1=\sum_{j<N_E}|\widehat F_j-F_j|(E_{j+1}-E_j)$, where $N_E$ is the number of energy coordinates.

\clearpage

\subsection{Structural response under sparse family adaptation}\label{note:f3}
\subsubsection{Strain-response fidelity across the four compounds}
The improvement in strain-response prediction extends across all four compounds and both strain signs. Shared adaptation reduced absolute-spectrum and response N-L2 for every withheld state, with adapted response N-L2 ranging from 0.184 to 0.312 and response PCC from 0.950 to 0.983 (Supplementary Table~\ref{tab:si-strain-cases}). The signed $Q$-integrated profiles in Supplementary Fig.~\ref{fig:si-strain-family} complement the WSe$_2$ maps in Fig.~\ref{fig:3}a by resolving the energy-dependent intensity gains and losses under both tension and compression.

\begin{figure}[!htbp]
\centering
\includegraphics[width=\textwidth,height=0.65\textheight,keepaspectratio]{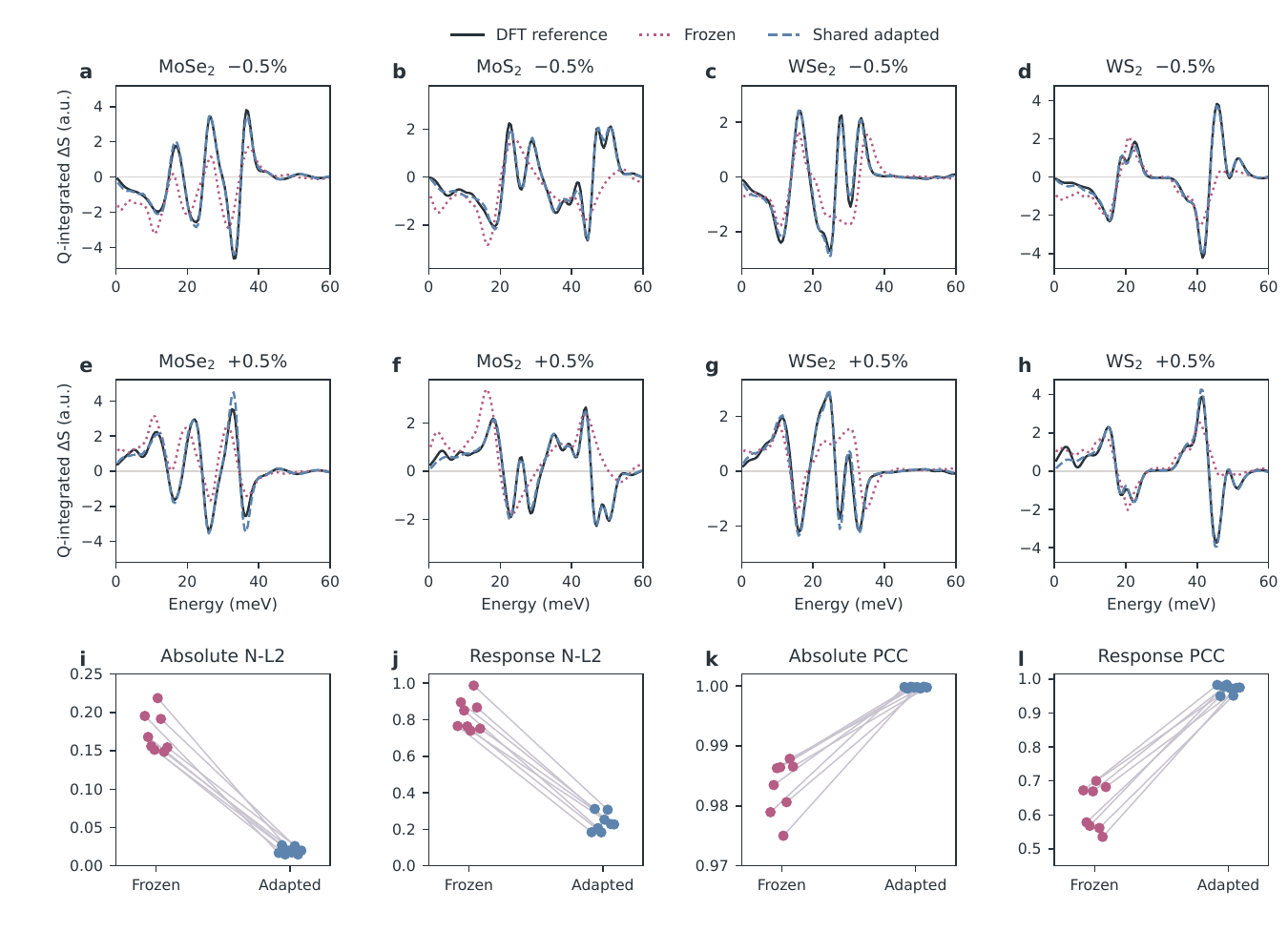}
\caption{\textbf{Strain-response prediction across four compounds.} \textbf{a--h,} $Q$-integrated difference spectra at withheld $-0.5\%$ (top) and $+0.5\%$ (bottom) strains, each calculated relative to the corresponding unstrained spectrum. Frozen denotes the global model before adaptation; Shared adapted denotes the native-grid strain model. Both strain signs share a vertical scale within each material; profiles display 0--60~meV. \textbf{i--l,} Full-grid absolute-spectrum and response N-L2 and PCC for all eight states (Supplementary Table~\ref{tab:si-strain-cases}), with segments connecting the global and adapted results.}
\label{fig:si-strain-family}
\end{figure}

\begin{table*}[htbp]
\centering\small
\caption{\textbf{Full-grid fidelity for all eight withheld strain states.} G: global model before family-specific adaptation; A: shared strain-adapted model. Absolute-spectrum and finite-response metrics use the complete $300\times299$ grid.}
\label{tab:si-strain-cases}
\begin{tabular}{lrrrrrrr}
\hline
Material & $\varepsilon$ (\%) & Abs. N-L2 G & Abs. N-L2 A & $\Delta S$ N-L2 G & $\Delta S$ N-L2 A & $\Delta S$ PCC G & $\Delta S$ PCC A\\\hline
MoSe$_2$ & $-0.5$ & .1954 & .0169 & .7653 & .1844 & .6724 & .9828\\
MoSe$_2$ & $+0.5$ & .2186 & .0175 & .7400 & .2532 & .7002 & .9736\\
MoS$_2$ & $-0.5$ & .1680 & .0270 & .8952 & .3117 & .5783 & .9500\\
MoS$_2$ & $+0.5$ & .1915 & .0259 & .9869 & .3074 & .5616 & .9514\\
WSe$_2$ & $-0.5$ & .1559 & .0149 & .8503 & .2067 & .5681 & .9784\\
WSe$_2$ & $+0.5$ & .1488 & .0149 & .8670 & .2289 & .5355 & .9741\\
WS$_2$ & $-0.5$ & .1512 & .0200 & .7633 & .1838 & .6695 & .9834\\
WS$_2$ & $+0.5$ & .1544 & .0200 & .7511 & .2272 & .6823 & .9752\\\hline
Mean & & .1730 & .0196 & .8274 & .2379 & .6210 & .9711\\\hline
\end{tabular}
\end{table*}

\subsubsection{Withheld compositions and organization of material representations}
The 30 withheld configurations include 18 interior and 12 boundary compositions within the sampled composition rectangle. Median N-L2 decreased from 0.4111 (95\% interval $[0.3811,0.4422]$) before adaptation to 0.0924 ($[0.0617,0.1139]$) after adaptation, with a 95\% paired-bootstrap interval of $[-0.359,-0.283]$ for the adapted-minus-global difference. The resampling procedure is specified in \hyperref[sec:si-f3-statistics]{Composition statistics and representation analysis}.

A complementary comparison examines the spectral changes between adjacent compositions. Across all 76 adjacent pairs in the 45-point composition grid, including training and withheld compositions, median spectral-response N-L2 decreased from 0.981 to 0.616 after adaptation. This improvement extends the absolute-spectrum result to the intensity changes along the substitution series.

To examine whether the composition ordering in Fig.~\ref{fig:3}c also reflects local spectral dissimilarity between adjacent compositions, we compared representation and reference-spectral distances along these same links. Spearman correlations were $-0.205$ before adaptation and $-0.022$ afterwards, with 95\% intervals of $[-0.445,0.026]$ and $[-0.259,0.214]$, respectively. After adaptation, directional correlations were $-0.300$ along W substitution and $0.383$ along Se substitution. The ordered substitution trends therefore describe compositional organization, while distances between adjacent representations do not provide a uniform measure of the associated spectral change.

\subsubsection{Transfer to independently generated larger configurations}
Reconstruction errors showed no monotonic increase with configuration size at either the training composition or the withheld composition (Supplementary Table~\ref{tab:si-size-transfer}). Across all ten independently randomized and relaxed configurations, full-grid N-L2 remained between 0.0595 and 0.0938, extending the 300-atom example in Fig.~\ref{fig:3}d across the 12--300-atom size series.

\begin{table*}[htbp]
\centering\small
\caption{\textbf{Size-transfer fidelity of the composition-adapted model.} Full-grid comparison with MACE--Phonopy--OCLIMAX references for independently randomized and relaxed configurations, without further training. S/U denote compositions included/withheld during adaptation, respectively.}
\label{tab:si-size-transfer}
\begin{tabular}{rrrrrr}\hline
$n\times n$ & Atoms & N-L2 S & PCC S & N-L2 U & PCC U\\\hline
$2\times2$ & 12 & .0845 & .9957 & .0938 & .9947\\
$4\times4$ & 48 & .0595 & .9979 & .0686 & .9972\\
$6\times6$ & 108 & .0909 & .9950 & .0710 & .9970\\
$8\times8$ & 192 & .0800 & .9962 & .0855 & .9956\\
$10\times10$ & 300 & .0765 & .9965 & .0725 & .9968\\\hline
\end{tabular}
\end{table*}

\subsubsection{Computational capacity beyond reference-validated sizes}
The 11,532-atom inference reached a peak allocated memory of 91.64~GiB during neighbor-graph construction. At the next tested size, 12,288 atoms, graph construction exhausted the available GPU memory. The limiting stage at these sizes was therefore construction of the atomic neighborhood graph, before evaluation of the spectral coordinates.

\subsubsection{Calculation and analysis details}\label{sec:si-f3-details}
\paragraph{Reference spectra.}
Strain-reference calculations used PBE with D3 dispersion corrections~\cite{perdew1996pbe,grimme2010d3} and SSSP 1.3 Precision pseudopotentials~\cite{prandini2018sssp}. Wavefunction and charge-density cutoffs were 60 and 480~Ry, respectively, with an electronic convergence threshold of $10^{-10}$~Ry. Unshifted electronic meshes were $12\times12\times4$ for unit-cell relaxation and $3\times3\times4$ for displaced supercells. Fixed-cell BFGS relaxation used force and energy thresholds of $10^{-4}$~Ry/Bohr and $10^{-5}$~Ry. Phonopy~\cite{togo2023phonopy} used $4\times4\times1$ force supercells, 0.01~\AA\ displacements and a Gamma-centered $32\times32\times8$ phonon mesh, with force-constant symmetrization enabled and without non-analytical corrections.

The specialized MACE potential~\cite{siddiqui2024alloys} supplied fixed-cell BFGS relaxation to 0.005~eV/\AA\ and double-precision finite-displacement forces. Phonopy used $4\times4\times1$ replicas, 0.01~\AA\ displacements, disabled symmetry and a Gamma-centered $32\times32\times1$ phonon mesh. Both reference datasets used OCLIMAX~\cite{cheng2019oclimax} on the native $300\times299$ powder grid under the temperature and resolution conditions specified in \hyperref[sm:data]{Methods}, with \texttt{TASK=1}, \texttt{INSTR=3}, \texttt{MAXO=10} and \texttt{CONV=2}. The phonon-energy cutoff \texttt{ECUT} was 0.01~meV for native-grid strain references and 0.1~meV for the native-grid composition and size-transfer references. Normalization setting 1 in OCLIMAX 3.0 divided intensities by the summed atomic scattering cross sections.

The separate strain-readout targets used case-dependent INS Gaussian broadening parameters of three source energy bins and two source momentum bins. The spectra were linearly resampled onto the common domain listed in Supplementary Table~\ref{tab:si-protocol-map}, whose bounds were defined from the overlapping training and validation support.

\paragraph{Adaptation and checkpoint selection.}\label{sec:si-f3-adaptation}
Adaptation followed the training and selection protocols in \hyperref[sm:adaptation]{Methods}. For strain checkpoints tied in mean validation response N-L2, selection favored higher response PCC and then the earlier epoch.

\paragraph{Composition statistics and representation analysis.}\label{sec:si-f3-statistics}
Composition-fidelity intervals used 2,000 configuration-bootstrap resamples, with identical draws for paired global--adapted comparisons, including the difference in medians. For the descriptive PCA projection specified in \hyperref[sm:geometry]{Methods}, PC1 and PC2 explained 67.6\% and 29.9\% of variance.

Linkwise distances used one minus cosine similarity in the original, uncentered representation space and the shape-sensitive reference-INS distance defined in \hyperref[sm:geometry]{Methods}. Correlation intervals used 2,000 source-composition bootstrap resamples, retaining each source's linked pairs as a block.

The full grid contains 76 adjacent pairs: 36 W-substitution steps of $\Delta x=0.25$ and 40 Se-substitution steps of $\Delta y=0.125$. For each pair $(i,j)$, spectral-response N-L2 compared $\widehat S_j-\widehat S_i$ with $S_j-S_i$ over the complete reference grid, normalized by $\max(\|S_j-S_i\|_2,10^{-12})$. The reported median includes all 76 pairs, not only pairs involving withheld compositions; these finite differences were not divided by the composition increment.

\paragraph{Larger-configuration references and capacity measurements.}
For lateral sizes $n=2,4,6,8,10$, in-plane force-supercell replication factors were 2, 1, 1, 1 and 1, and phonon meshes were $32\times32\times1$, $16\times16\times1$, $11\times11\times1$, $8\times8\times1$ and $7\times7\times1$, respectively. No non-analytical or explicit acoustic-sum-rule correction was applied. Sensitivity to the force-supercell size was examined for the 12-atom midpoint configuration: increasing the in-plane replication factor from 2 to 4 changed the reference spectrum by an N-L2 of 0.0193.

The 11,532-atom benchmark used expandable CUDA allocation segments with the FP32 model and hardware specified in \hyperref[sm:timing]{Methods}. Peak memory was measured using PyTorch allocator statistics and CUDA device queries in separate profiling runs.

\par\bigskip

\subsection{Observable-specific neighborhoods, targeted analysis and computational cost}\label{note:f4}
\subsubsection{Observable supervision selectively reorganizes neighborhoods}
Matching spectroscopic supervision improves both the average reference-spectral similarity of learned neighborhoods and recovery of the nearest reference-spectral neighbors across the 649 test materials (Supplementary Table~\ref{tab:si-representation-geometry}). The paired 95\% query-bootstrap intervals for enrichment gains over initialization are $[0.074,0.095]$ for INS and $[0.031,0.041]$ for eDOS.

\begin{table}[htbp]
\centering\small
\caption{\textbf{Reference-spectral similarity of learned neighborhoods.} Enrichment and neighbor recall for the matched models in Fig.~\ref{fig:4}, evaluated on 649 test materials. Larger enrichment $1-R(k)$ indicates closer spectral neighbors; zero denotes random-neighbor similarity. Brackets give 95\% query-material bootstrap intervals. Recall@5 is the overlap fraction between representation-selected and reference-spectral neighbor sets (random expectation 0.0077).}
\label{tab:si-representation-geometry}
\begin{tabular}{llrrr}
\toprule
Representation & Observable & $1-R(5)$ [95\% interval] & Recall@5 & $1-R(20)$ \\
\midrule
Initialization & INS & 0.361 [0.346, 0.377] & 0.179 & 0.226 \\
Initialization & eDOS & 0.157 [0.147, 0.167] & 0.152 & 0.095 \\
INS-only & INS & 0.445 [0.430, 0.460] & 0.230 & 0.371 \\
INS-only & eDOS & 0.147 [0.137, 0.158] & 0.144 & 0.093 \\
eDOS-only & INS & 0.363 [0.347, 0.379] & 0.171 & 0.264 \\
eDOS-only & eDOS & 0.193 [0.182, 0.204] & 0.204 & 0.149 \\
Joint & INS & 0.387 [0.371, 0.402] & 0.181 & 0.301 \\
Joint & eDOS & 0.191 [0.180, 0.201] & 0.192 & 0.146 \\
\bottomrule
\end{tabular}
\end{table}

Matching supervision increases Recall@5 from 0.179 to 0.230 for INS and from 0.152 to 0.204 for eDOS. Joint supervision instead leaves INS recall nearly unchanged (0.181 versus 0.179), despite higher enrichment: its neighbors are closer in mean INS distance without a comparable increase in exact-neighbor overlap. Matching-observable enrichment gains also persist at $k=10,20$ (Supplementary Fig.~\ref{fig:si-neighbor-transitions}a,b).

Neighbor replacement identifies how the supervised response gains priority within the learned neighborhoods. Among queries whose neighbor sets changed, the gained neighbors were closer than the lost neighbors in the supervised observable, with 95\% intervals for gained-minus-lost distance of $[-0.109,-0.087]$ under INS supervision and $[-0.076,-0.058]$ under eDOS supervision. Evaluating the same replacements against the other observable gave more distant eDOS spectra under INS supervision, whereas the change in INS distance under eDOS supervision was not resolved (Supplementary Fig.~\ref{fig:si-neighbor-transitions}c,d).

\begin{figure}[!htbp]
\centering
\includegraphics[width=\textwidth,height=0.61\textheight,keepaspectratio]{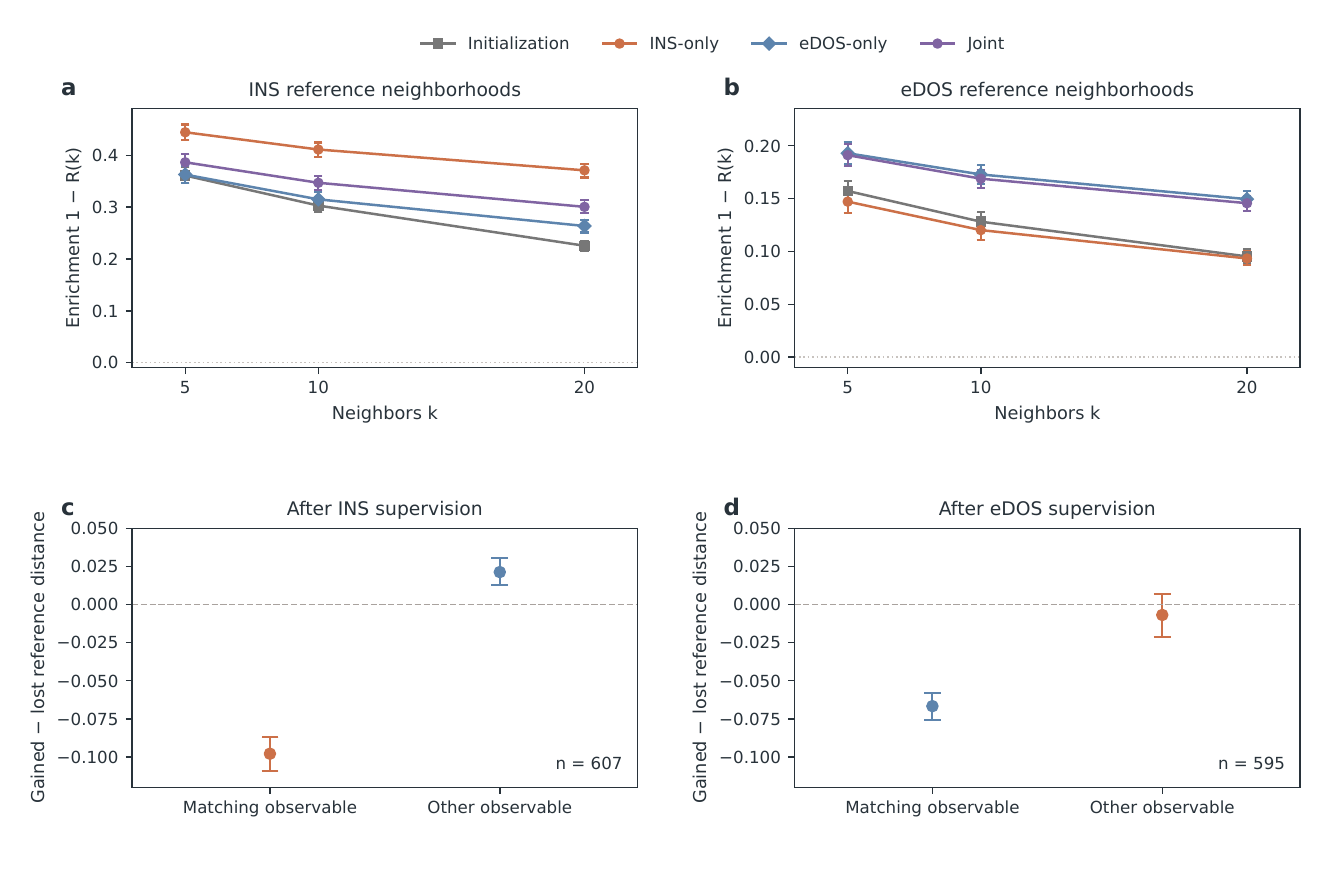}
\caption{\textbf{Observable-dependent enrichment and neighbor replacement.} Results use the matched models shown in Fig.~\ref{fig:4}. \textbf{a,b,} Enrichment $1-R(k)$ for 649 test materials at $k=5,10,20$. Larger values indicate closer reference-spectral neighbors; zero is the random-neighbor expectation. Bars are 95\% query-material bootstrap intervals. \textbf{c,d,} Mean gained-minus-lost reference distance for queries whose neighbor sets change: 607 under INS and 595 under eDOS supervision. Negative values indicate closer replacements. Matching- and cross-observable distances are shown.}
\label{fig:si-neighbor-transitions}
\end{figure}

\paragraph{Dependence on training seed and neighborhood size.}
Matching-observable enrichment gains, their advantage over cross-observable supervision, and joint-supervision gains for both observables persisted across all three training seeds and $k=5,10,20$, with paired query-bootstrap intervals above zero. Cross-observable effects depended on neighborhood size: INS supervision reduced eDOS enrichment at $k=5$ in every seed, whereas eDOS supervision gave INS-gain intervals spanning zero at $k=5$ and positive, smaller gains at $k=10,20$. The preference for the supervised response thus persisted even where both spectral similarities improved. Seed-resolved results and across-seed standard deviations are reported in Supplementary Table~\ref{tab:si-seed-robustness}.

Within each seed, both INS and eDOS coordinate networks were instantiated in the same order for every supervision condition, preserving the matched initialization specified in \hyperref[sm:matched-training]{Methods}. Only active-task losses contributed gradients.

\begin{table}[!htbp]
\centering\small
\setlength{\tabcolsep}{4pt}
\caption{\textbf{Training-seed robustness of spectral-neighborhood enrichment.} Results at $k=5$ for 649 test materials. Seeds 1, 2 and 3 are 20260729, 20260730 and 20260731, respectively. Own-initialization gains are $\Delta\mathcal{E}(5)=\mathcal{E}_{\mathrm{trained}}(5)-\mathcal{E}_{\mathrm{own\ init.}}(5)$. Matching-minus-cross contrasts compare INS-only with eDOS-only supervision within INS neighborhoods and eDOS-only with INS-only supervision within eDOS neighborhoods. Values are dimensionless; $\pm$ denotes the sample standard deviation across three training seeds.}
\label{tab:si-seed-robustness}
\begin{tabular}{llrrrr}
\toprule
Contrast & Observable & Seed 1 & Seed 2 & Seed 3 & Mean $\pm$ SD \\
\midrule
INS-only $-$ own init. & INS & $0.08353$ & $0.07335$ & $0.07658$ & $0.07782 \pm 0.00520$ \\
INS-only $-$ own init. & eDOS & $-0.01006$ & $-0.01017$ & $-0.01239$ & $-0.01087 \pm 0.00132$ \\
eDOS-only $-$ own init. & INS & $0.00180$ & $-0.00371$ & $0.00040$ & $-0.00050 \pm 0.00287$ \\
eDOS-only $-$ own init. & eDOS & $0.03571$ & $0.03612$ & $0.03518$ & $0.03567 \pm 0.00047$ \\
Joint $-$ own init. & INS & $0.02553$ & $0.02401$ & $0.02535$ & $0.02496 \pm 0.00083$ \\
Joint $-$ own init. & eDOS & $0.03388$ & $0.03415$ & $0.03225$ & $0.03343 \pm 0.00102$ \\
\midrule
Matching $-$ cross & INS & $0.08173$ & $0.07706$ & $0.07618$ & $0.07832 \pm 0.00298$ \\
Matching $-$ cross & eDOS & $0.04577$ & $0.04628$ & $0.04757$ & $0.04654 \pm 0.00093$ \\
\bottomrule
\end{tabular}
\end{table}

\subsubsection{Sampling accuracy of local spectral quantities}
To quantify the error introduced by limited coordinate sampling, we compared sparse estimates of the local quantities with estimates from the complete local grid of the same model. Across 30 withheld compositions, 90 direct queries per structure gave centroid and upper-band-fraction MAEs of 0.043~meV (95\% interval $[0.029,0.058]$) and 0.4186 percentage points, respectively (Supplementary Table~\ref{tab:si-query-budgets}). Increasing the budget to 1,717 queries reduced these sampling errors to 0.0051~meV and 0.0152 percentage points.

\begin{table}[htbp]
\centering\small
\caption{\textbf{Direct-query errors across 30 withheld compositions.} Query fractions use the full 89,700-point grid. Upper-band-fraction errors are expressed in percentage points (pp). Full-model errors against the MACE reference are 0.1236~meV for the centroid and 1.1572~pp for the upper-band fraction.}
\label{tab:si-query-budgets}
\begin{tabular}{rrrrrr}
\toprule
Points & Grid (\%) & \multicolumn{2}{c}{Centroid MAE (meV)} & \multicolumn{2}{c}{Weight MAE (pp)} \\
 & & vs model & vs reference & vs model & vs reference \\
\midrule
30 & 0.0334 & 1.0479 & 0.9826 & 2.2211 & 1.7821 \\
90 & 0.1003 & 0.0430 & 0.1082 & 0.4186 & 1.0713 \\
238 & 0.2653 & 0.0556 & 0.1183 & 0.1286 & 1.0885 \\
594 & 0.6622 & 0.0205 & 0.1147 & 0.0717 & 1.1125 \\
1717 & 1.9142 & 0.0051 & 0.1214 & 0.0152 & 1.1497 \\
\bottomrule
\end{tabular}
\end{table}

The centroid sampling error did not decrease monotonically with query count, being larger at 238 than at 90 queries (Supplementary Fig.~\ref{fig:si-composition-queries}). Applying the same 90-coordinate quadrature to the reference spectra gave a centroid error of 0.045~meV, comparable to the 0.043~meV error for the learned field. Thus, the small numerical integration error at this budget was also obtained when sampling the reference spectrum directly.

Errors against the reference combine the sampling error within the predicted field and the full-grid prediction error. Their signed contributions can partially cancel, giving a smaller centroid MAE at 90 queries than with complete-local-grid evaluation (0.108 versus 0.124~meV), with the same ordering for the upper-band fraction. Consequently, convergence toward the full-model estimate and agreement with the reference follow different curves in Supplementary Fig.~\ref{fig:si-composition-queries}.

\begin{figure}[!htbp]
\centering
\includegraphics[width=\textwidth,height=0.42\textheight,keepaspectratio]{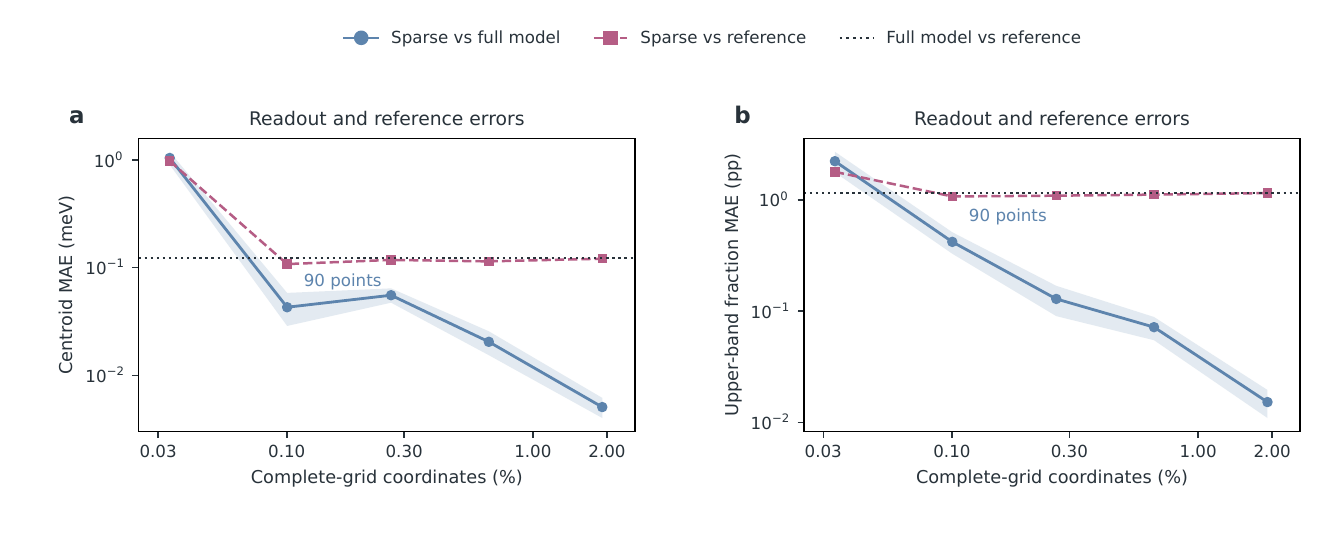}
\caption{\textbf{Sampling accuracy of local spectral quantities.} \textbf{a,b,} Centroid and upper-band-fraction MAEs versus direct-query budget for 30 withheld compositions: blue, sparse versus full-model estimates; rose, sparse estimates versus MACE references; dotted, full-model estimates versus references. Shading gives 95\% configuration-bootstrap intervals for the blue curves. Query percentages use the full 89,700-point grid; both axes are logarithmic. Upper-band-fraction errors are expressed in percentage points.}
\label{fig:si-composition-queries}
\end{figure}

\paragraph{Direct-query evaluation.}
The energy-window endpoints and threshold were aligned to the nearest native energy coordinates, approximately 9.9997, 59.9979 and 29.9990~meV, respectively. All budgets used the same 30 withheld compositions and unchanged composition-adapted model. Confidence intervals used 2,000 configuration-bootstrap resamples.

\subsubsection{Ridge shifts and spectral-weight redistribution describe different responses}
Stored predictions from the strain model trained on broadened and resampled targets (Supplementary Table~\ref{tab:si-protocol-map}) gave a ridge-energy MAE of 0.049~meV across the eight withheld states at $\pm0.5\%$ strain. The four finite-difference coefficients recovered the reference ordering of response magnitudes, with an MAE of 0.044~meV/\% and PCC of 0.911. The coefficient error was largest for MoS$_2$ (0.132~meV/\%), compared with 0.008--0.022~meV/\% for the other compounds (Fig.~\ref{fig:4}e).

Separating intensity gain and loss before momentum integration retains the redistribution that can cancel in the signed energy profile, as illustrated for MoSe$_2$ in Supplementary Fig.~\ref{fig:si-random-local}a. Across the eight withheld states, integrated intensity gains and losses were recovered with MAEs of 0.0131 and 0.00981, respectively, in the arbitrary units of the integrated reference spectra (Supplementary Fig.~\ref{fig:si-random-local}b,c).

\begin{figure}[!htbp]
\centering
\includegraphics[width=\textwidth,height=0.40\textheight,keepaspectratio]{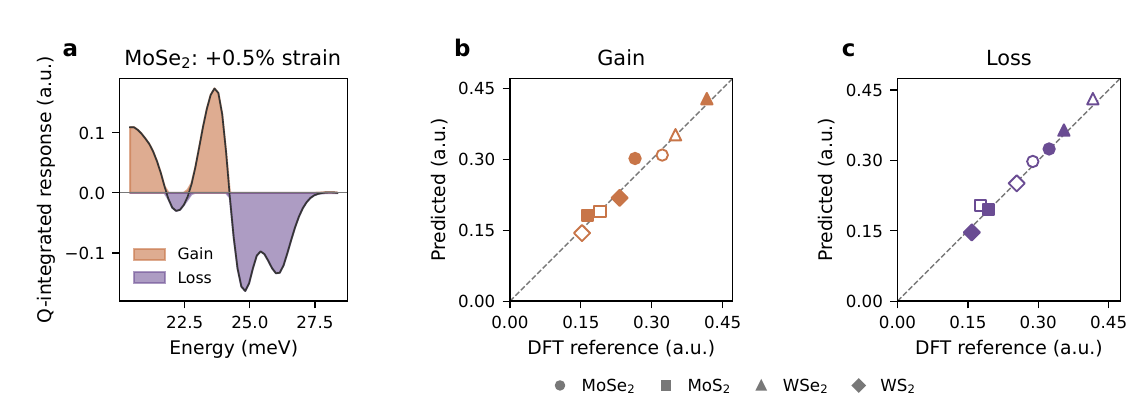}
\caption{\textbf{Local spectral-weight redistribution under strain.} \textbf{a,} Positive and negative parts of the MoSe$_2$ reference response at $+0.5\%$ strain, separated before $Q$ integration; the black line is their signed sum. \textbf{b,c,} Gain and loss integrated over fixed local windows, comparing strain-readout predictions with DFT references for all four materials at $\pm0.5\%$ strain ($n=8$). Shapes identify materials; filled and open markers denote positive and negative strain. Dashed lines indicate parity.}
\label{fig:si-random-local}
\end{figure}

\paragraph{Coefficient sampling controls.}
Subsampling stored predictions at zero strain and the validation $\pm1.5\%$ states isolated numerical sampling effects from the withheld-state comparison with DFT. With ten $Q$ samples, increasing the number of energies from five to seventeen reduced the mean deviation from the complete-local-grid coefficient from 0.0348 to 0.00235~meV/\%, using 0.0556\% and 0.1889\% of the full coordinates across the three states, respectively. Twenty-five energies reduced the deviation further to 0.000312~meV/\% at a sampling fraction of 0.2778\%.

Low sampling errors were also obtained with randomly selected energy samples from the stored spectra. At the same 17-energy budget, regular sampling gave a mean coefficient deviation of 0.00235~meV/\%. For random sampling, the corresponding four-material mean deviations had a median of 0.00328~meV/\% across 200 realizations. The 95\% paired hierarchical-bootstrap interval for random-minus-regular deviation, $[-0.00185,0.00769]$~meV/\%, did not distinguish the two strategies at this budget.

Both strategies used the same window width, $Q$ integration, tracking rule and peak estimator. Random coordinates were drawn without replacement. Each of 10,000 bootstrap replicates resampled materials and then a random realization for each selected material.

\paragraph{Ridge selection.}
Candidate ridges were selected using the training and validation reference spectra at $Q$-window centers of 2, 4, 6 and 8~\AA$^{-1}$. Zero-strain candidates had prominence of at least 3\% of the profile maximum and a minimum separation of eight energy bins. Candidates were tracked in both strain directions by selecting the nearest local maximum within 2~meV of the preceding peak, with intensity breaking distance ties. Fully tracked candidates were ranked by $p(|s|+0.1)\max(R^2,0)$, where $p$ is zero-strain prominence, $s$ is the numerical slope of a linear fit to ridge energy versus strain in meV per percentage point, and $R^2$ describes that fit. The highest-scoring ridge for each material occurred at $Q=8$~\AA$^{-1}$. The corresponding zero-strain anchors $E_{0,m}$ were approximately 24.38, 20.69, 26.44 and 17.60~meV for MoSe$_2$, MoS$_2$, WSe$_2$ and WS$_2$, respectively.

\paragraph{Peak interpolation and local integration.}
For peak interpolation, the quadratic was fitted through the sampled maximum and its two neighbors at their actual energy coordinates, which need not be equally spaced. A concave fit with its vertex between the neighboring samples supplied the peak estimate; boundary maxima or invalid fits retained the sampled maximum. This estimator was used for reference and predicted spectra, at every budget and for both regular and random sampling.

Each fixed gain/loss window $\Omega_m$ contained 55 energy coordinates and ten $Q$ coordinates on the resampled strain-readout grid. Equation~\ref{eq:local-gain-loss} in \hyperref[sm:queries]{Methods} was evaluated using trapezoidal weights determined by the physical coordinate spacings, retaining the reference intensity scale.

\clearpage
\subsubsection{Scattering post-processing dominates large-cell reference-generation cost}
Stage-resolved timings explain why accelerating force evaluation alone leaves a substantial cost for powder-INS generation. Under the fixed alloy protocol, OCLIMAX accounted for more than 99.9\% of the recorded reference-generation time at 300 atoms (Supplementary Table~\ref{tab:si-timing-contracts}). The dominant cost therefore occurred after the MACE force and phonon calculations, whereas SpecSCoNe evaluated the complete spectrum directly. Reconstruction fidelity for the same size series is reported in Supplementary Table~\ref{tab:si-size-transfer}.

\paragraph{Execution settings.}
Each WSe$_2$ DFT force job used one RTX PRO 6000 Blackwell GPU with one MPI process and one OpenMP thread. Both WSe$_2$ learned routes used the same GPU model and four CPU threads. Post-processing used a $32\times32\times8$ phonon mesh without non-analytical correction and zero-temperature OCLIMAX calculations, with normalization setting 1, maximum order 10 and resolution parameters of approximately 0.456~meV and 0.0625~\AA$^{-1}$. OCLIMAX executed on a single CPU thread. For the alloy reference series, one spectrum-generation run was recorded per configuration; the launcher set the OpenMP and numerical-library thread limits to 16 and allowed up to ten concurrent pipelines. The size-dependent phonon meshes and force-supercell choices are listed in Supplementary Note~\ref{note:f3}.

\begin{table}[!htbp]
\centering\small
\caption{\textbf{Stage-resolved spectrum-generation times (seconds).} \textbf{A,} DFT, MLFF and SpecSCoNe routes for the same relaxed WSe$_2$ structure on the $300\times300$ strain-analysis domain. DFT force time is summed job time; FC denotes force-constant construction; MLFF time combines the median force/force-constant time with one post-processing run. \textbf{B,} Alloy-size comparison on the native $300\times299$ domain, with medians over two configurations per size. Phonopy/convert includes phonon-mesh evaluation and OCLIMAX-input conversion. SpecSCoNe timings include graph construction, encoding and complete-grid evaluation; final CPU transfer is included in A and excluded in B. Dashes denote stages bypassed by direct prediction. Measurement protocols are specified in \hyperref[sm:timing]{Methods}.}
\label{tab:si-timing-contracts}
\begin{tabular}{lrrr}\toprule
\multicolumn{4}{l}{\textbf{A. Single-structure comparison}}\\
Route & Force jobs / FC & Post-processing & Complete spectrum\\\midrule
DFT & 15,600 & 9.074 & $\simeq1.56\times10^4$\\
EquiformerV3 MLFF & 0.366 & 9.055 & 9.421\\
SpecSCoNe & --- & --- & 0.0217\\\bottomrule
\end{tabular}

\medskip
\begin{tabular}{rrrrrr}\toprule
\multicolumn{6}{l}{\textbf{B. Alloy size series}}\\
Atoms & MACE force/FC & Phonopy/convert & OCLIMAX & Total reference & SpecSCoNe\\\midrule
12 & 4.279 & 8.450 & 1,186 & 1,199 & 0.0323\\
48 & 5.714 & 30.52 & 18,582 & 18,618 & 0.0367\\
108 & 14.23 & 72.87 & 98,031 & 98,118 & 0.0474\\
192 & 35.04 & 125.88 & 288,846 & 289,007 & 0.0645\\
300 & 75.00 & 247.08 & 843,666 & 843,989 & 0.0836\\\bottomrule
\end{tabular}
\end{table}

\subsubsection{Phonon-mesh and scattering-approximation controls}\label{sec:si-oclimax-protocol}
The cost of scattering post-processing can be reduced by coarser phonon sampling or by changing the scattering approximation, with different consequences for the calculated intensity distribution. We examined these effects separately for the same relaxed 300-atom configuration at $(x,y)=(0.5,0.5)$, reusing its MACE force constants (Supplementary Fig.~\ref{fig:si-oclimax-protocol}).

Reducing the \texttt{TASK=1} mesh from $7\times7\times1$ to $2\times2\times1$ lowered recorded OCLIMAX time from 234.3 to 19.48~h, with a full-grid N-L2 difference of 0.0314 relative to the original calculation. The principal bands remained similar, while 97.9\% of the squared difference occurred below 5~meV, identifying the low-energy response as the main mesh-sensitive region.

At the original $7\times7\times1$ mesh, the incoherent approximation (\texttt{TASK=0}) required 72.7~s and gave full-grid N-L2 difference 0.0685. The difference is smaller for the $Q$-integrated energy profile (0.00729) but larger at fixed $Q=2$~\AA$^{-1}$ (0.470), showing that momentum integration can conceal local sensitivity to the scattering approximation.

\paragraph{Control calculation settings.}
The controls used the common spectral settings and OCLIMAX-only timing protocol described in \hyperref[sm:timing]{Methods}. With \texttt{MAXO=10} and \texttt{CONV=2}, second- and higher-order contributions were evaluated by convolution. Numerical-library thread limits were unchanged; the launcher permitted up to ten concurrent pipelines in the original workflow and two in the controls.

\begin{figure}[!htbp]
\centering
\includegraphics[width=\textwidth,height=0.60\textheight,keepaspectratio]{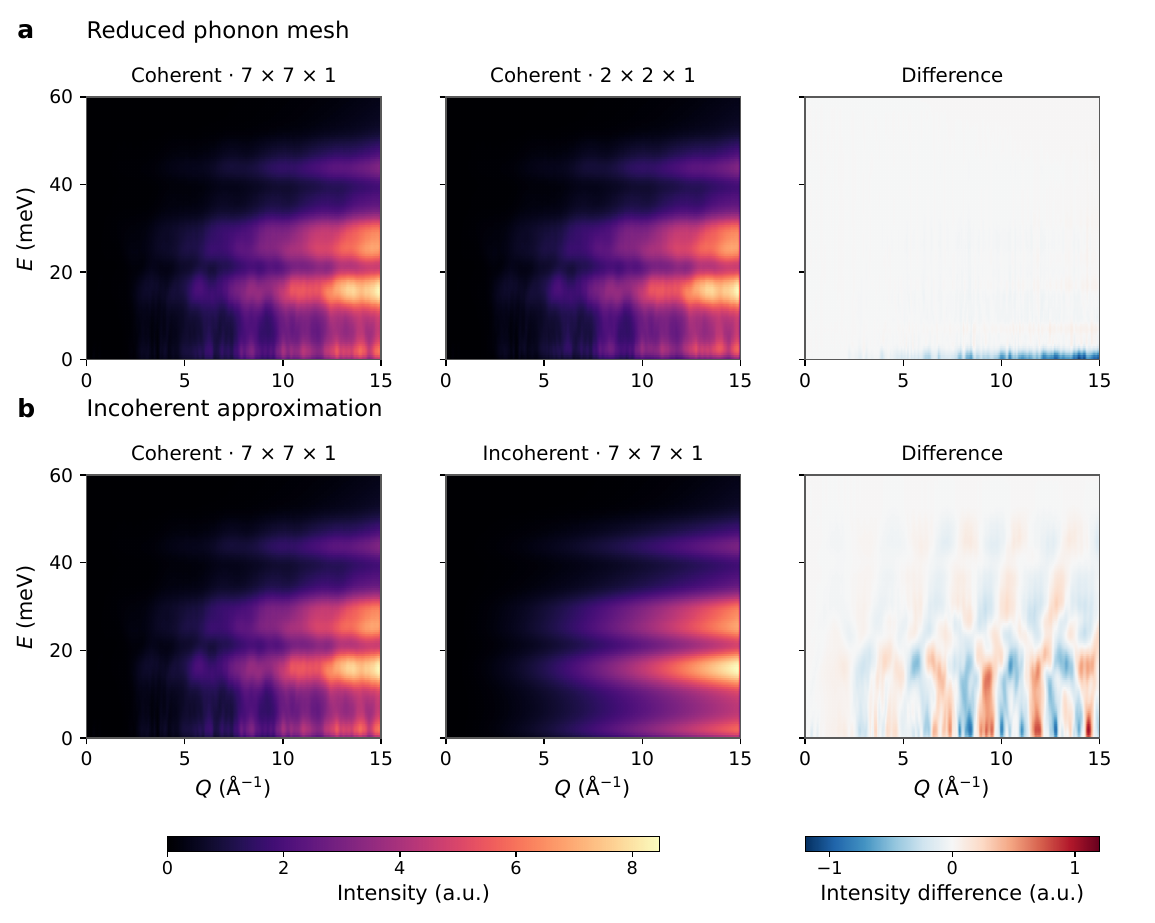}
\caption{\textbf{Phonon-mesh reduction and scattering-approximation controls.} Powder-INS spectra for the same relaxed 300-atom alloy configuration at $(x,y)=(0.5,0.5)$. \textbf{a,} Coherent calculations with $7\times7\times1$ and $2\times2\times1$ phonon meshes. \textbf{b,} Coherent and incoherent-approximation calculations with the same $7\times7\times1$ mesh. Coherent denotes OCLIMAX \texttt{TASK=1}, including coherent and incoherent contributions; the incoherent approximation uses \texttt{TASK=0}. The left column repeats the original coherent calculation, and the right column shows the middle spectrum minus the left spectrum. Absolute spectra share an intensity scale, and both difference maps share a symmetric scale. Maps display 0--60~meV; numerical comparisons use the complete native $300\times299$ energy--momentum grid without additional intensity rescaling.}
\label{fig:si-oclimax-protocol}
\end{figure}

\clearpage
\endgroup